\documentclass[twocolumn,pre,showpacs]{revtex4}
\usepackage{epsfig}
\usepackage{graphicx}
\usepackage{epsfig}
\usepackage{dcolumn}
\usepackage{bm}
\newcommand{\drv}{\textrm{d}}
\begin{document}

\title{Lamellar order, microphase structures and glassy phase
in a field theoretic model for charged colloids}
\author{Marco Tarzia$^{a}$ and Antonio Coniglio$^{a,b}$}
\affiliation{${}^a$ Dipartimento di Scienze Fisiche and INFN sezione di Napoli,
Universit\`{a} degli Studi di Napoli ``Federico II'',
Complesso Universitario di Monte Sant'Angelo, via Cinthia, 80126 Napoli, Italy}
\affiliation{${}^b$ Coherentia CNR-INFM}
\date{\today}

\begin{abstract}
In this paper we present a detailed analytical study of the phase diagram and
of the structural properties of a field theoretic model with a short-range
attraction and a competing long-range screened repulsion.
We provide a full derivation and expanded discussion and digression on results 
previously reported briefly in 
M. Tarzia and A. Coniglio, Phys. Rev. Lett. {\bf 96}, 075702
(2006).
The model contains the essential features of the effective
interaction potential among charged colloids in
polymeric solutions.
We employ the self-consistent Hartree approximation
and a replica approach, and 
we show that varying the parameters of the repulsive potential and the
temperature yields a phase coexistence, a lamellar and a glassy phase.
Our results
suggest that the cluster phase observed in charged colloids might
be the signature of an underlying equilibrium lamellar phase, hidden on 
experimental time scales, and emphasize that the formation of microphase
structures may play a prominent 
role in the process of colloidal gelation.
\end{abstract}
\maketitle

\section{Introduction}\label{sec:intro} 
Colloidal suspensions are solutions of solid (or 
liquid) mesoscopic particles immersed into another substance~\cite{coll_books}.
These systems, like blood, proteins in water, milk, inks, or paints, 
are ubiquitous in our everyday life and are extremely important in biology and 
industry.
Due to their potential applications 
for designing new materials with a wide range of viscoelastic properties,
in the last few years there has been much interest in the
role of the inter-particle potential on controlling the structural and
dynamical properties of colloidal systems.
By appropriately varying some control parameters (such as the composition of 
the solvent, the coating of the particles, the concentration of the 
polymers into the solvent, \ldots) the effective interaction potential between
colloids can be suitably tuned in the experiments. It is possible to realize
a hard-sphere system \cite{pusey}; by adding non adsorbing polymers the 
hard-sphere interaction can be  complemented by a short-range attraction, 
induced by depletion forces~\cite{lekk}. 
Recent experimental works outlined that in some cases a residual net charge  
on the surface of colloidal particles may be present~\cite{strad,campbell}, 
thereby inducing a 
long-range electrostatic repulsion screened by the presence of ions in the
solution. The resulting effective interaction is therefore given by a 
hard-core term accounting for the excluded volume, a depletion-induced 
narrow attractive shell 
and a long-range repulsive shoulder. This kind of potential is well 
approximated by the DLVO (after Derjaguin, Landau, Verwey and Overbeek) 
potential~\cite{dlvo}.
Interestingly enough, the parameters of the potential can be suitably tuned 
in the experiments. 
The depth of the attractive shell is controlled by 
the concentration of the polymers, $\phi_p$, which therefore plays the 
role of an inverse temperature, whereas its range is proportional
to the ratio between the radius of the colloids, 
$\sigma$, and the gyration 
radius of the polymers, $R_g$ (typically one has $0.05 < R_g/ 
\sigma < 0.2$). 
On the other
hand, the addition of salt increases the number of ions in the 
solution responsible for the screening of the electrostatic repulsion, thereby 
reducing the amplitude and increasing the screening length 
of the Yukawa potential.

Charged colloids in polymeric solutions have 
recently raised a lot of interest, both from an 
experimental~\cite{strad,campbell,weitz} and a 
theoretical~\cite{sciort_wig,sciortino,sator,char_reich,reatto,letter,tubi} 
point of view. 
In these systems the
competition between attractive and repulsive interactions on
different length scales stabilizes the formation of aggregates of an
optimal size and shape ({\em cluster phase}), characterized by a peak of the
structure factor around a typical wave vector, $k_m$.
Experimentally, such cluster phase, made up by approximately monodisperse 
equilibrium aggregates, can be clearly observed  
using confocal microscopy at low volume fraction and 
low temperature (high attraction strength)~\cite{strad,weitz,campbell}.
By appropriately tuning the control parameters (i.e., increasing the
volume fraction or decreasing the temperature) the system
progressively evolves toward a gel-like non-ergodic disordered 
state~\cite{weitz,campbell} 
({\em colloidal gelation}), where structural arrest occurs. Although 
intensely studied both experimentally and numerically, a theoretical
understanding these phenomena is still lacking: the mechanisms inducing
colloidal gelation are still unknown and many gaps remain in our
present knowledge of the equilibrium phase diagram of these systems.
In Ref.~\cite{sciort_wig}, it has been proposed that 
colloidal gelation is related to the formation of a Wigner glass,
whose blocks are compact and thermodynamically stable clusters, with a
residual long-range repulsion between them.
In Ref.~\cite{sator} it was instead highlighted the percolative nature of the 
structural arrest in charged colloids, 
suggesting that percolation of a spanning network of 
clusters with long living bonds plays a crucial role.
More recently, it has been suggested that the formation of partially ordered 
anisotropic domains and microphase structures may be responsible for physical
gelation in these systems~\cite{char_reich,letter,tubi}.

In a recent letter~\cite{letter}, we have introduced
a $\phi^4$ model with
competition between a short-range attraction, described by the
Ginzburg-Landau Hamiltonian, and a long-range screened repulsion,
described by a Yukawa potential~\cite{lattice}. Albeit schematically, the model 
contains the essential features of the 
effective interaction potential among charged colloids in polymeric solutions
and sheds new light on the structural properties of these systems.
Here we present a detailed and extensive analytical study of this 
model, and a full derivation and explanation of the results previously reported
briefly in~\cite{letter}.

We show that depending on the control parameters, there is a
region of the phase diagram where usual phase separation between a 
colloid-rich and a colloid-poor phase takes
place. Conversely, as the screening length and/or the strength of
the repulsion exceeds a threshold value, phase separation is
prevented. In this case, at moderately high temperature the
competition
between attraction and repulsion has the effect to produce microphase 
structures, i.e., partially ordered modulated domains 
which in the terminology of particle systems, correspond
to the cluster phase. These modulated structures are the precursors
of a first order transition to an equilibrium lamellar phase found
at lower temperatures.
By using a replica approach for systems without quenched
disorder~\cite{monasson,mezard}, and by employing the Self-Consistent Screening
Approximation (SCSA)~\cite{bray,mezard1}, we also show the presence of a
glass transition line in the low temperature region, once the first
order transition to the lamellar phase is avoided. The
mechanism responsible for the glass transition in this case turns out
to be completely
different from that of molecular glass-formers: as a matter of fact, we find 
that glassiness is not due to the
presence of a hard-core interaction, which is totally absent in our model;
it is instead due to the
formation of the microphase structures which order up to the size of
correlation length.
The geometric frustration, arising from arranging such modulated
structures in a disordered fashion, leads to a complex free energy
landscape and, consequently, to a dynamical slowing down.

Our results suggest that the cluster phase observed in
colloidal suspensions should be followed, upon decreasing the
temperature (or increasing the volume fraction), by an equilibrium
periodic phase (a columnar or a lamellar phase, depending on the
volume fraction). If, instead, this ordered phase is avoided, a structural
arrest, corresponding to the gel phase observed in
the experiments and in numerical simulations,
should eventually occur.
%
The existence of ordered phases in colloidal suspensions and the fact that  
the transition to the gel phase could occur in a metastable liquid, are novel 
predictions, which have never been considered before.
Recently, 
the presence of such ordered phases in atomistic model
systems of charged colloids interacting via the DLVO potential, has been 
unambiguously shown both in three~\cite{char_reich,tubi} and in 
two~\cite{char_reich,reatto} dimensions.

The paper is organized as follows:
in Sec.~\ref{sec:model} we describe
the model; in Sec.~\ref{sec:hart} we study the model	
within the self-consistent Hartree approximation; in
Sec.~\ref{sec:dyn} we analyze the dynamics of the system 
in the spherical limit; 
in Sec.~\ref{sec:glass} we study the
glass transition; in Sec.~\ref{sec:length} we compute the
pair correlation function of the system in the real space,
showing the emergence of competing
length scales and emphasize the prominent role of microphase
structures; in Sec.~\ref{sec:phdi} we summarize the results found:
we discuss the resulting phase diagram
and stress the connections with charged colloids.
Some details of the calculations are reported in 
Apps.~\ref{app:hart}-\ref{app:scsa}.

\section{The model} \label{sec:model}
We consider the standard three
dimensional $\phi^4$ field-theory with the addition of a repulsive
long-range potential~\cite{letter}:
\begin{equation} \label{eq:rs.model}
{\cal H} [\phi] = \int \drv^3 \mathbf{x} \left[ f(\phi) + \frac{1}{2}
\left(\nabla \phi (\mathbf{x})\right)^2\right] + \mathcal{H}_{LR} [\phi],
\end{equation}
where $\phi (\mathbf{x})$ is the scalar order parameter field, related to
the concentration of colloidal particles. 
The local free energy has the usual Ginzburg-Landau form:
\begin{equation}
f(\phi) = \frac{r_0}{2} \, \left( \phi (\mathbf{x}) \right)^2 + \frac{g}{4} \,
\left( \phi (\mathbf{x}) \right)^4.
\end{equation}
The coefficient $r_0$ is a temperature dependent mass proportional to the
deviation $r_0 \propto T - T_c^{MF}$ from the mean field transition temperature
in absence of frustration. The long-range repulsive interaction is described by
a Yukawa potential:
\begin{equation}
\mathcal{H}_{LR} [\phi ] = \frac{W}{2} \int \!\!\!\! \int \drv^3 \mathbf{x} \,
\drv^3 \mathbf{x}^{\prime} \, \frac{e^{- | \mathbf{x} - \mathbf{x}^{\prime}|/
\lambda}} {|\mathbf{x} - \mathbf{x}^{\prime}|} \, \phi(\mathbf{x})
\phi(\mathbf{x}^{\prime}).
\end{equation}
The local term favours the formation of a uniform condensed phase, while
the long-range term energetically frustrates this condensation.
In the momentum space the Hamiltonian of Eq.~(\ref{eq:rs.model}) reads:
\begin{eqnarray} \label{eq:coll.model}
&& {\cal H} [ \phi ] =
\frac{V}{2} \int \frac{\drv^3 {\bf k}}{(2 \pi)^3} \, \left[
r_0 + k^2 + \frac{4 \pi W}{\lambda^{-2} + k^2} \right] \, \phi_{\bf k}
\phi_{- {\bf k}}\\
\nonumber
&& \,\,\, + \, \frac{g V}{4}
\int \frac{\drv^3 {\bf k_1}}{(2 \pi)^3} \, \frac{\drv^3 {\bf k_2}}{(2 \pi)^3}
\, \frac{\drv^3 {\bf k_3}}{(2 \pi)^3} \,
\phi_{\bf k_1} \phi_{\bf k_2} \phi_{\bf k_3} \phi_{- {\bf k_1} - {\bf k_2} -
{\bf k_3}}.
\end{eqnarray}
The model has been also studied in~\cite{westhfal} in the context of
microemulsion. 
The parameters $W$ and $\lambda$ are, respectively,
the strength and the range of the repulsive potential. For $W=0$ we
obtain the canonical short-range ferromagnet. Interestingly, for
$\lambda \to \infty$ we recover the case of Coulomb
interaction. 
The latter model was first introduced in
Ref.~\cite{emery}, in the context of cuprate systems. Then, it has been
further studied in many other papers (see, for instance,
Refs.~\cite{coniglio,kivelson,tarjus,wolynes,wolynes1,tarjus1})
where it has been used to describe the phenomenology of a wide
variety of systems, where competing interactions on different length
scales stabilize pattern formation and the creation of spatial
inhomogeneities (for a review see~\cite{review}). These systems
include magnetic systems  and dipolar fluids characterized by
long-range Coulombic interactions~\cite{roland}, mixtures of block
copolymers~\cite{ohta}, water-oil-surfactant
mixtures~\cite{stillinger} and doped Mott insulator, including 
high $T_c$ superconductors~\cite{tranquada}. 
As a consequence, our
model allows to describe and to interpret in an unified fashion the
phenomenology of a broad range of different physical systems.

\section{Paramagnetic, Ferromagnetic and Lamellar phases within the
Hartree approximation} \label{sec:hart}
\begin{figure}
\begin{center}
\includegraphics[scale=0.31,angle=270]{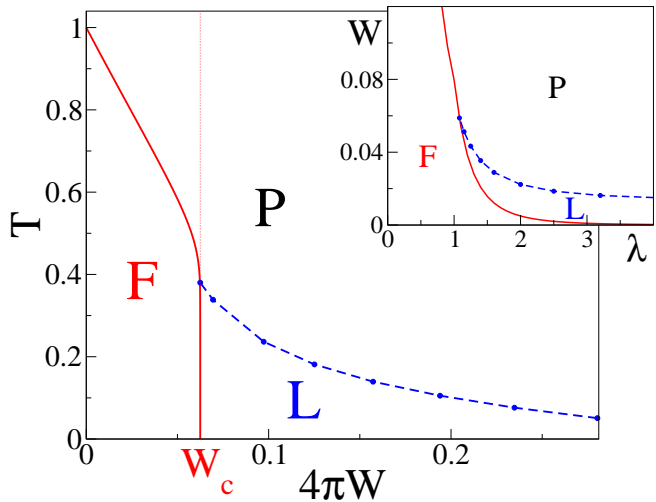}
\end{center}
\caption{(color online) {\bf Main frame:}
Phase diagram of the model obtained within the self-consistent Hartree
approximation, in the frustration ($4 \pi W$)-temperature ($T$) plane, for a
fixed range of the repulsive potential, $\lambda=2$ (and $g=1$, $r_0 = -1$).
The temperature is rescaled with respect to the critical temperature of the
unfrustrated model, $T_c (W=0) = 2 \pi^2/3 \Lambda$.
The (red) continuous curve, $T_c (W, \lambda)$, corresponds to
the second order phase transition from the paramagnetic to the ferromagnetic
phase (with the usual Hartree critical exponents, $\nu = 1$ and $\gamma=2$),
whereas the (blue) dashed curve corresponds to the fluctuation-induced first
order transition from the paramagnetic to the lamellar phase,
$T_L (W, \lambda)$. $W_c$ is the critical threshold of the repulsion
strength,
Eq.~(\ref{eq:coll.wc}). 
5are set in such a way that the elementary length of a lattice spacing, is
{\bf Inset:} System phase diagram in the range ($\lambda$)-frustration ($W$)
plane for a fixed temperature $T=0.1$, showing the relative position of the
different phases. 
}
\label{fig:coll.hart}
\end{figure}
In this section we solve the model, Eq.~(\ref{eq:rs.model}), 
within the self-consistent Hartree approximation.
This approximation amounts in
replacing one factor $\phi^2$ in the quartic term by its average, namely
$\langle \phi^2 \rangle$, to be determined self-consistently~\cite{ChLu}.
For a one component order parameter there are six ways of choosing the two
factor of $\phi$ to be paired in $\langle \phi^2 \rangle$ among the four
factor of $\phi$ in the quartic term $g \phi^4 /4$. Hence, the self-consistent
Hartree approximation consists in substituting the term $g \phi^4 /4$
with $3 g \langle \phi^2 \rangle \phi^2 /2$. With this substitution, the
Hamiltonian becomes quadratic:
\begin{equation}
Z = \int {\cal D} \phi_{\bf k} \, e^{- \frac{\beta V}{2} \! \int \!
\frac{d^3 {\bf k}}{(2 \pi)^3} \, \left[
r_0 + 3 g \langle \phi^2 \rangle + k^2 +
\frac{4 \pi W}{\lambda^{-2} + k^2} \right] \phi_{\bf k}
\phi_{- {\bf k}}}.
\end{equation}
It is then possible to compute $Z$ and 
evaluate the correlation function:
\begin{equation}
G ( {\bf k} ) = \langle \phi_{\bf k} \phi_{-{\bf k}} \rangle - \langle
\phi_{\bf k} \rangle \langle \phi_{-{\bf k}} \rangle.
\end{equation}
In the paramagnetic phase, $\langle \phi_{\bf k} \rangle = 0$, 
one obtains:
\begin{eqnarray} \label{eq:coll.gk}
\nonumber
G ( {\bf k} ) &=& \frac{T}{r_0 + 3 g \langle \phi^2 \rangle + k^2 +
\frac{4 \pi W}{\lambda^{-2} + k^2}} \\
&=& \frac{T}{r + k^2 + \frac{4 \pi W}{\lambda^{-2} + k^2}},
\end{eqnarray}
where the renormalized mass term, $r$, has been defined as:
\begin{equation} \label{eq:coll.scha1}
r \equiv r_0 + 3 g \langle \phi^2 \rangle.
\end{equation}
We now recall that:
\begin{equation} \label{eq:coll.scha2}
\langle \phi^2 \rangle = G ( {\bf x}, {\bf x}) = \int_{|{\bf k}| < \Lambda}
\frac{\drv^3 {\bf k}}{(2 \pi)^3} \, G ({\bf k}),
\end{equation}
where $\Lambda$ is the ultraviolet cutoff.
As a result, Eqs.~(\ref{eq:coll.scha1}) and (\ref{eq:coll.scha2}) yield:
\begin{equation} \label{eq:coll.scha}
r = r_0 + 3 g \int_{|{\bf k}| < \Lambda}
\frac{\drv^3 {\bf k}}{(2 \pi)^3} \, \frac{T}{r + k^2 + \frac{4 \pi
W}{\lambda^{-2} + k^2}},
\end{equation}
which provides a self-consistent equation for the renormalized mass $r$.
It can be shown that, apart from the factor $3$, this
approximation becomes exact for an $N$-component order parameter in the 
limit $N\to \infty$~\cite{ChLu}.

In order to solve the self-consistent equation, Eq.~(\ref{eq:coll.scha}), it is
convenient to define:
\begin{equation} \label{eq:coll.wc}
W_c \equiv \frac{1}{4 \pi \lambda^4},
\end{equation}
and distinguish between two cases: $W<W_c$ and $W>W_c$.
\subsection{$W<W_c$}
For $W<W_c$ the correlation function, $G ({\bf k})$,
behaves as in the standard unfrustrated case: it is a
monotonically decreasing function of $| {\bf k}|$,
with a maximum in $k=0$, where it equals:
\begin{equation}
G ({\bf k} = {\bf 0}) = \frac{T}{r + 4 \pi W \lambda^2}.
\end{equation}
Therefore, as $r \to - 4 \pi W \lambda^2$, the susceptibility,
$\chi = G ({\bf k} = {\bf 0}) / T$, diverges. This corresponds to a usual
second order phase transition towards a ferromagnetic phase. From
Eq.~(\ref{eq:coll.scha}), 
one can compute the critical
temperature, $T_c (W, \lambda)$, given by the following relation:
\begin{equation}
\frac{3 g T_c}{2 \pi^2} \int_0^{\Lambda} \drv k \, \frac{1 +
k^2}{\left(1 + \frac{W}{W_c} \right) + k^2} = -r_0 - 4 \pi W \lambda^2.
\end{equation}
It is possible to show that the usual Hartree critical exponents are found,
i.e., $\nu = 1$ and $\gamma=2$ (see App.~\ref{app:hart}).
In the terminology of colloidal suspensions, this means that for $T<T_c$ the
system undergoes a phase separation between a colloid-rich and a colloid-poor
phase.
The dependence of the critical temperature, $T_c (W, \lambda)$, upon
$W$ (for $\lambda = 2$ and $r_0 = -1$) is plotted in the main frame of
Fig.~\ref{fig:coll.hart} (red continuous line in Fig.~\ref{fig:coll.hart}),
showing that the only effect of the repulsive
interaction for $W<W_c$ is to decrease the numerical value of $T_c$: 
for $W = 0$ we recover the
usual critical temperature of the unfrustrated system in the Hartree
approximation, 
$T_c (W = 0) = - \frac{2 \pi^2 r_0}{3 g \Lambda}$
(notice that $r_0 < 0$), whereas $T_c$ vanishes as $W \to W_c$.

\subsection{$W>W_c$}
Conversely, if $W > W_c$, the correlation function has a maximum around a
finite value of the wave vector, $k_m$:
\begin{equation} \label{eq:coll.kmax}
k_m = \left( \sqrt{4 \pi W} - \lambda^{-2} \right)^{1/2} = \left ( 4 \pi
\right)^{1/4} \left(
\sqrt{W} - \sqrt{W_c} \right)^{1/2}.
\end{equation}
As we shall discuss later on, the peak in the correlator indicates that 
the system establishes microphase structures with 
inverse domain size given by $k_m$, which characterize 
the incipient periodic order. Such modulated structures are the analog
of the cluster phase observed in colloidal systems characterized, also in
this case, by a peak in the structure factor around a characteristic
wave vector corresponding to the inverse of the typical size of the 
clusters~\cite{campbell,weitz,sciort_wig,sciortino,sator}.

In $k=k_m$, the correlation function is given by
\begin{equation}
G ( | {\bf k} | = k_m ) = \frac{T}{r + 2 \sqrt{4 \pi W} - \lambda^{-2}}.
\end{equation}
Therefore, $G(|{\bf k}| = k_m)$ and, consequently, the fluctuations, $\chi
(|{\bf k}| = k_m)$, diverge for $r \to r_{sp} \equiv \lambda^{-2} - 2 \sqrt{4
\pi W}$. As a result, also the integral of the right hand side of
Eq.~(\ref{eq:coll.scha}) diverges, and the self-consistent Hartree relation
can be satisfied only at $T =0$.
This implies that for $W>W_c$ the paramagnetic phase is stable at all
finite temperatures, its spinodal line, where the susceptibility for
$|{\bf k}| = k_m$ diverges, $\chi (k_m) \to \infty$, is located
at $T = 0$. 
Therefore above $W_c$
there is no phase separation. This result is quite important for
designing new materials as well as in
the experimental and numerical study
of colloidal systems, where it is crucial to distinguish the slowing
down due to colloidal gelation from that due to kinetic of phase
separation.

Nevertheless, the system still undergoes
a first order transition to a lamellar phase 
when the temperature is lowered
below $T_L (W, \lambda)$
(blue dashed line in Fig.~\ref{fig:coll.hart}).
Such first order phase
transition 
is induced by the
fluctuations, as first discussed by Brazovskii for a related
model
\cite{brazov} and as also found in the Coluombic case
($\lambda \to \infty$)~\cite{tarjus1}.
The transition temperature $T_L (W, \lambda)$ can be determined on equating 
the free energy of the paramagnetic phase to that of the lamellar 
phase~\cite{tarjus1}, as explained in App.~\ref{app:fifot}. 

The lamellar phase is characterized by a periodic variation of the
order parameter with wave-length $l_m = 2 \pi k_m^{-1}$:
\begin{equation} \label{eq:coll.lamel}
\langle \phi_{\bf k} \rangle = m \left( \delta \left( {\bf k} - {\bf k}_m
\right) + \delta \left( {\bf k} + {\bf k}_m \right) \right),
\end{equation}
where $k_m$ is given in Eq.~(\ref{eq:coll.kmax}).
Note that as $W$ approaches $W_c$ from above, according to
Eq.~(\ref{eq:coll.kmax}), $k_m$ vanishes. This signals the fact that
the size of the stripes diverges on the boundary between the lamellar
and the ferromagnetic phase.

Within the fluctuation approach described in App.~\ref{app:fifot}, it 
is also possible to study the stability of other kinds of ordered phases,
such as columnar or periodically ordered cluster phases, which occur
in numerical simulations~\cite{reatto,tubi}. 
In fact, it turns out that these phases
can be stable besides the lamellar phase, at lower volume fractions,
i.e., for $\langle \phi_{{\bf k} = {\bf 0}} \rangle < 0$. However, in this 
paper we only focus on the case $\langle \phi_{{\bf k} = {\bf 0}} \rangle 
= 0$, (i.e., $\int \textrm{d}^3 {\bf x} \langle \phi ({\bf x}) \rangle = 0$)
where the only periodically ordered stable phase is the lamellar one.
 
\subsection{Phase diagram within the Hartree approximation}
The results found within the self-consistent Hartree approximation are
summarized in Fig.~\ref{fig:coll.hart}, where the
system phase diagram in the frustration ($4 \pi W$)-temperature ($T$) plane at
fixed screening length $\lambda = 2$ (main frame) and in
the screening length 
($\lambda$)-frustration ($W$) plane at fixed temperature $T=0.1$
(inset), is plotted, showing the relative positions of the paramagnetic (P), 
the ferromagnetic (F) and the lamellar (L) phases.

According to Eq.~(\ref{eq:coll.wc}), by changing $\lambda$, $W_c$ changes and,
hence, the phase diagram modifies. For instance, in
the Coulomb limit, $\lambda \to \infty$, we have that $W_c \to 0$.
As a consequence, the ferromagnetic phase reduces to the axis $W = 0$, for $T <
T_c (W=0)$. Conversely, for $\lambda \to 0$ we have that $W_c \to \infty$ and
the lamellar phase disappears.
It is interesting to observe that the dependence of the threshold $W_c$
on the screening length $\lambda$ 
is exactly the same of that found in numerical
simulations of a model system
for charged colloids interacting via a DLVO potential~\cite{sciortino}:
the authors find that for $W < W_c$ the system tends to phase separate,
whereas for $W>W_c$     
the preferred structures are elongated (and eventually quasi
one-dimensional) finite clusters.
Physically, such threshold value, $W_c$,
can be interpreted 
expanding the Hamiltonian, Eq.~(\ref{eq:coll.model}), 
up to the lowest order in $k^2$:
\begin{eqnarray} \label{eq:coll.exp}
&& {\cal H} [\phi] \simeq 
\frac{V}{2} \int \frac{\drv^3 {\bf k}}{(2 \pi)^3} \Big[
\left( r_0 + 4 \pi W \lambda^2 \right ) \\
\nonumber
&& \qquad \qquad \qquad \qquad \,\,\,\,\,\,\,
+ \left(1 - 4 \pi W \lambda^4\right) k^2 \Big] \phi_{\bf k}
\phi_{- {\bf k}}\\
\nonumber
&& \,\,
+ \, \frac{gV}{4}
\int \frac{\drv^3 {\bf k_1}}{(2 \pi)^3} \, \frac{\drv^3 {\bf k_2}}{(2 \pi)^3}
\, \frac{\drv^3 {\bf k_3}}{(2 \pi)^3} \, \phi_{\bf k_1}
\phi_{\bf k_2} \phi_{\bf k_3}
\phi_{ - {\bf k_1} - {\bf k_2} - {\bf k_3}}.
\end{eqnarray}
At $W = W_c$ the coefficient of the $k^2$ term vanishes, implying that 
there is no energetic cost associated to the creation
of interfaces between high density and low density regions, 
leading to pattern formation and to the creation of spatial inhomogeneities.

\section{Dynamics in the spherical limit} \label{sec:dyn}
In this section we study 
the dynamics of the model in the case of
an $N$-component order parameter in the limit $N\to \infty$,
after an instantaneous
quench from high temperature. 
The spherical approximation is equivalent to the Hartree approximation
described in the previous section (apart from the factor $3$ in
Eq.~(\ref{eq:coll.scha})), and
allows to study analytically the dynamics.
We solve the time dependent Cahan-Hilliard
equation~\cite{cah_hil},    
following the approach taken
in Refs.~\cite{con_glo}. We obtain an exact analytical
solution for the evolution of the time dependent structure factor of the
system,
$G(\mathbf{k},t)$, and for the peak position, $k_{max}(t)$.
In agreement with the results presented in the previous section, 
we find that as $W<W_c$, the
usual spinodal decomposition occurs, and the maximum
of the structure factor approaches zero as $t^{-1/4}$ for large times.
This corresponds to phase separation between a high density 
colloid-rich phase and low density colloid-poor one.
On the other hand, for $W>W_c$, provided that we are on the spinodal line of
the lamellar phase ($T=0$), the peak position approaches a non-zero
value, $k_m$, given in Eq.~(\ref{eq:coll.kmax}), and the structure factor
asymptotically approach a $\delta$-function peaked around $k_m$, signaling
a lamellar order. 
However, since the paramagnetic phase is stable at all finite temepratures,
for $T>0$ we find that
the lamellar phase starts to form locally, but then fades away 
after long time and the system stays homogeneous:
$k_{max}(t)$ approaches a
limiting value where the structure factor vanishes exponentially.

Let us consider the Cahan-Hilliard equation~\cite{cah_hil}:
\begin{equation}
\frac{\partial \phi (\mathbf{x}, t)}{\partial t} =
M \nabla^2 \frac{\delta {\cal H} [\phi (\mathbf{x}, t)]}
{\delta \phi (\mathbf{x}, t)},
\end{equation}
where
$M$ is related to the mobility, and ${\cal H}$ is given in
Eq.~(\ref{eq:rs.model}). For an $N$-component order parameter
the above equation reads:
\begin{eqnarray} \label{eq:coll.NTDGL}
\nonumber
\frac{\partial \phi_{\alpha} (\mathbf{x}, t)}{\partial t} &=&
M \nabla^2 \Bigg[ \Bigg( r + \frac{g}{N} \sum_{\beta = 1}^N \phi_{\beta}^2
(\mathbf{x}, t) - \nabla^2 \Bigg) \phi_{\alpha} (\mathbf{x}, t) \\ 
&& \,\,
+ \, W \int \drv^3 \mathbf{x}^{\prime} \frac{e^{- | \mathbf{x} - \mathbf{x'}|
/\lambda}
\phi_{\alpha}
(\mathbf{x}^{\prime},t)}{|\mathbf{x} - \mathbf{x}^{\prime}|} \Bigg].
\end{eqnarray}
In the limit $N\to \infty$ we recover the spherical model replacing
$\frac{1}{N} \sum_{\beta=1}^{N} \phi^2_{\beta} (\mathbf{x}, t)$ by
$\langle \phi^2_{\alpha} (\mathbf{x}, t)\rangle$.
Here $\langle \cdot \rangle$
represents an ensemble average over the initial configurations.   
By assuming the translational invariance of the pair correlation
function $G(\mathbf{x}, \mathbf{x}^{\prime}, t) =
\langle \phi (\mathbf{x}, t) \phi (\mathbf{x}^{\prime}, t) \rangle
= G(|\mathbf{x} - \mathbf{x}^{\prime}|, t)$, the quantity
$\langle \phi^2 (\mathbf{x}, t)\rangle = G(0,t) \equiv S(t)$ becomes
independent on the position $\mathbf{x}$.
Thus, dropping the index
$\alpha$, taking the Fourier transform
over the space of Eq.~(\ref{eq:coll.NTDGL}) and multiplying both sides by
$\phi_{\alpha} (-{\bf k},t)$
we obtain the equation of motion for the time dependent
structure factor, $G (\mathbf{k}, t) = \langle \phi_{\alpha} ({\bf k},t)
\phi_{\alpha} (-{\bf k},t) \rangle$:
\begin{equation} \label{eq:coll.SK}
\frac{\partial G (\mathbf{k}, t)}{\partial t} =
- M k^2 \left[ r + g S(t) + k^2 + \frac{4\pi W}{\lambda^{-2} + k^2} \right]
G (\mathbf{k}, t).
\end{equation}
By replacing $\frac{1}{N} \sum_{\beta=1}^{N} \phi^2_{\beta} (\mathbf{x}, t)$
by $S(t)$ we have, in fact, enabled a linearization of
Eq.~(\ref{eq:coll.NTDGL}), ``preaveraging'' the nonlinear term.
Now Eq.~(\ref{eq:coll.SK}) can be readily integrated to obtain:
\begin{equation} \label{eq:coll.solSK}
G (\mathbf{k}, t) = G (\mathbf{k}, 0) \, e^{- M k^2 \left[
Q(t) + k^2t + \frac{4\pi W t}{\lambda^{-2} + k^2} \right]},
\end{equation}
where $Q(t)$ is defined by
\begin{equation} \label{eq:Qt}
Q(t) = \int_0^t \drv t^{\prime} \left[ g S(t^{\prime}) + r \right].
\end{equation}
To complete the
solution, $S(t)$ must be computed self-consistently. 
As shown in App.~\ref{app:dynamics}, one can obtain an expression
for the time dependent structure factor, Eq.~(\ref{eq:coll.Gkt}), 
and an equation for the
evolution of the peak position:
\begin{eqnarray} \label{eq:coll.equation}
&& \exp \left[ M t \left( 1 -\frac{4 \pi W}
{(\lambda^{-2} + k_{max}^2(t))^2} \right) k_{max}^4(t)
\right] \\
\nonumber
&& \qquad \,\, = \, P \times
\left[ 1 - \frac{4 \pi W \lambda^{-2}}
{(\lambda^{-2} + k_{max}^2(t))^3}\right]^{1/2}
\frac{\sqrt{Mt}}{k_{max}(t)},
\end{eqnarray}
where the function $P$ is given in Eq.~(\ref{eq:P}).
Taking the logarithm
of both sides of Eq.~(\ref{eq:coll.equation}) and dividing by $Mt$ we obtain
an asymptotic expression
for $k_{max}(t)$, Eq.~(\ref{eq:coll.asympt}), 
which is readily analyzed in the limit $t \to \infty$.
Careful analysis show that
depending on the values of $W$ and $\lambda$,
different solutions of this equation are found. In particular, one
needs again to distinguish between the cases $W<W_c$ and $W>W_c$.

\subsection{$W<W_c$}
For $W < W_{c}$
the solution of Eqs.~(\ref{eq:coll.equation}) and (\ref{eq:coll.asympt}) is
given by:
\begin{equation}
k_{\infty} \equiv \lim_{t \to \infty} k_{max} (t) = 0,
\end{equation}
corresponding to the usual spinodal decomposition.
Using Eq.~(\ref{eq:coll.asympt}) we find that
that the usual $N \to \infty$ domain growth exponent
of the canonical short-range ferromagnet with conserved order parameter
is found:
\begin{equation}
k_{max}(t) \sim t^{-1/4}.
\end{equation}
Obviously, this solution
can be obtained provided that we are below the critical point,
$r < - 4 \pi W \lambda_0^2$ [$T<T_c(W)$], otherwise
no spinodal decomposition takes place
and the system remains in the paramagnetic phase.

\subsection{$W>W_c$}
For $W>W_{c}$ the
solution of Eqs.~(\ref{eq:coll.equation}) and (\ref{eq:coll.asympt}) 
is given by:
\begin{equation} \label{eq:coll.klamellar}
k^2_{\infty} = \sqrt{4\pi W} - \lambda^{-2} = k_m^2,
\end{equation}
where the wave vector $k_{\infty}$ coincides with
that defined in Eq.~(\ref{eq:coll.kmax}).
In this case, the peak position reaches a nonzero steady state value
for large times and the structure factor asymptotically approaches a
$\delta$-function peaked around a finite value of the wave vector $k_{m}$.
This situation
corresponds to the lamellar phase and $k_{m}$ is the inverse domain size.
In particular,
for earlier times (such that $4 \pi W \ll k_{max}^4(t)$), using
Eq.~(\ref{eq:coll.asympt}), we have that:
\begin{equation}
k_{max}(t) \sim \left( \frac{3}{4} \ln t \right)^{1/4} \, t^{-1/4}.
\end{equation}
Thus, apart from the logarithmic factor, in this regime
$k_{max}(t)$ behaves as the standard solution of the unfrustrated
short-range ferromagnet in the large
$N$ limit in the
case of usual spinodal decomposition. However, at later times,
$k_{max}(t)$
saturates at $k_m$ as a power law:
\begin{equation} \label{eq:approach}
k_{max}^2(t) \sim k_m^2 + C \, t^{-1/2}.
\end{equation}
According to Eq.~(\ref{eq:coll.Gkt}),  
the structure factor $G(\mathbf{k},t)$ can be
approximated by a Gaussian centred about $k_m$ and with width
\begin{equation}
\Delta k \sim \frac{(4 \pi W)^{1/4}}{\left[ \sqrt{4 M t}
(\sqrt{4 \pi W} - \lambda^{-2}) \right]},
\end{equation}
which in the limit $t \to \infty$ approaches a $\delta$-function
centred in $k_m$, as stated before.

However, this situation can occur only provided that we are on the spinodal 
line of the lamellar phase, located at $T=0$. As discussed in the previous 
section, the expression of the spinodal line reads
(both in the Hartree approximation and in the large $N$ limit):
\begin{equation} \label{eq:coll.rsp}
r_{sp} (W,\lambda) = - 2 \sqrt{4 \pi W} + \lambda^{-2}.
\end{equation}
Indeed, as $r > r_{sp} (W, \lambda)$ (i.e., $T>0$), from
Eq.~(\ref{eq:coll.asympt})
we have that $k_{max}(t)$
approaches exponentially
a limiting value, $k_{\infty}^2 = - (r + \lambda^{-2})/2$:
\begin{equation}
k_{max}^2(t) =
k_{\infty}^2 + C \times \frac{e^{-M (W - W_{sp}) t}}{\sqrt{Mt}},
\end{equation}
where $W_{sp} = (\lambda^{-2} - r)^2/16 \pi$.
Consequently, $P$, Eq.~(\ref{eq:P}),
asymptotically vanishes in an exponential fashion, signaling that 
the periodic pattern fades.
At the spinodal line we have that $W=W_{sp}$, which
gives back Eq.~(\ref{eq:approach}). 

In conclusion, the analysis of the dynamics in the spherical limit indicates 
that for $W<W_c$ the system undergoes a phase separation between a 
colloid-rich phase and a colloid-poor one. On the other hand, for
$W>W_c$ and $T>0$ the system starts to form microphase 
structures and periodically ordered modulated domains, 
as signaled by a peak in the structure factor which
grows in time and whose position seems to approach $k_m$. However,
since the paramagnetic phase is stable at all finite temperatures, the
microphase structures have a finite lifetime.
The lamellar order is fully established only at $T=0$.

\section{The Glass Transition}
\label{sec:glass}
\begin{figure}
\begin{center}
\includegraphics[scale=0.31,angle=270]{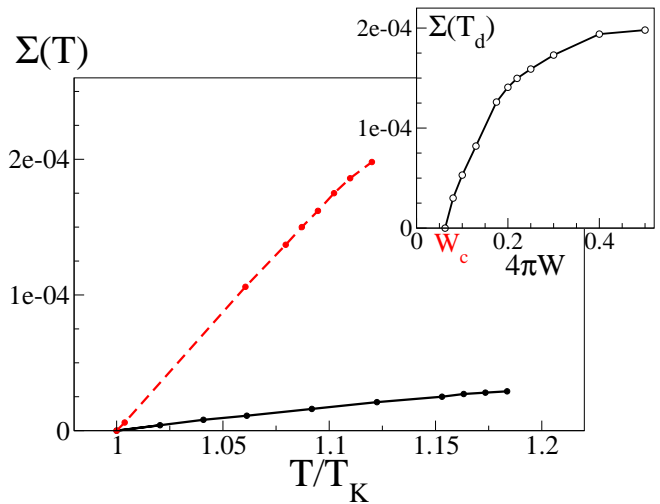}
\end{center}
\caption{(color online) 
{\bf Main frame:} Complexity, $\Sigma$, as a function of $T/T_K$ for $4 \pi
W = 0.5$ (red dashed curve) and $4 \pi W = 0.08$ (black continuous curve), for 
a fixed value of the range of the repulsive potential, $\lambda = 2$, and for
$g=1$ and $r_0 = -1$. At high temperature ($T>T_d$) the system is in the
paramagnetic phase and the complexity is zero. At $T=T_d$ the complexity
discontinuously jumps to a finite value, signaling the emergence of a
complex free energy landscape. The complexity decreases as the temperature is
decreased and vanishes at $T_K$ where the thermodynamic transition to a 1RSB 
glassy phase takes place. 
{\bf Inset:} Complexity at the dynamical transition temperature, $\Sigma 
(T_d)$, plotted as a function of the strength of the repulsive potential, $
4 \pi W$. As $W$ decreases, $\Sigma (T_d)$ decreases. At $W = W_c$, the glassy 
phase disappears. 
}
\label{fig:coll.sc}
\end{figure}
Since the paramagnetic phase is stable at all finite temperatures, the
first order transition to the lamellar phase can be kinetically avoided.
It is then possible to
supercool the system in the (meta-stable) homogeneous phase below the first 
order transition line, $T_L(W,\lambda)$, as much as  
one can obtain supercooled liquids by performing fast enough 
coolings below the melting temperature avoiding crystallization.
Experimental~\cite{ChBo92} and numerical~\cite{tarjus2,gonnella}
results show that, indeed, these
systems exhibit long-time relaxations similar to that observed in glasses.
Recent results
have clearly shown the presence of a glass transition in the
Coulomb case ($\lambda \to \infty$)~\cite{wolynes,wolynes1,tarjus1,reichman}.

\subsection{General formalism}
In order to study the glass transition in our model, we use a replica
approach introduced in Ref.~\cite{monasson} and then developed in
Ref.~\cite{mezard}, formulated to deal with system without quenched disorder,
which allows to compute the complexity, $\Sigma$.

Our aim is to derive the free energy landscape of the model.
The equilibrium free energy, defined as $F = - T \ln Z$,
is relevant only if the system is able to explore the entire phase space.
This is not the case in the glassy phase, where the system is frozen
in metastable states.
In order to scan the locally stable field configurations,
we introduce an appropriate symmetry breaking field $\psi
(\mathbf{x})$, and compute the following partition function~\cite{monasson}:
\begin{equation}\label{eq:coll.pinning}
\tilde Z [\psi] =  \int \mathcal{D} \phi \, \exp \left(
- \beta \mathcal{H} [\phi]
-\frac{u}{2} \int \drv^3 \mathbf{x} \left[ \psi (\mathbf{x}) -
\phi (\mathbf{x}) \right]^2 \right),
\end{equation}
where $u$ denotes the strength of the coupling. Since $u>0$, the introduction
of the pinning field $\psi (\mathbf{x})$ forces the order parameter $\phi
(\mathbf{x})$ to assume configurations close to $\psi (\mathbf{x})$.
Therefore, the free energy
\begin{equation}
\tilde f [\psi] = - T \ln \tilde Z [\psi]
\end{equation}
will be low if $\psi (\mathbf{x})$ equals to configurations
which locally minimize $\mathcal{H} [\phi]$. Thus, 
in order to scan all metastable states, we have to
sample all configurations
of the field $\psi$, weighted with $\exp(- \beta \tilde f [\psi])$:
\begin{equation}
\tilde F = \lim_{u \to 0^+} \frac{\int \mathcal{D} \psi \,
\tilde f [\psi] \, \exp \left( - \beta \tilde f [\psi] \right)}
{\int \mathcal{D} \psi \, \exp \left( - \beta \tilde f [\psi] \right)}.
\end{equation}
Therefore, $\tilde F$ 
is a weighted average of the free energy in the various metastable states.
If there are only a few local minima (i.e., a non-extensive number of local
minima) of free energy $\tilde f$ equal to $F$, we have that
$\tilde F$ equals the true free energy $F$ of the system.
However, in case of the emergence of an exponentially large number of
metastable states with large barriers between them, a nontrivial contribution
arises from the above integral even in the limit $u \to 0^+$ and $\tilde F$
differs from $F$. This allows to identify the configurational entropy
$\Sigma$~\cite{monasson,mezard}: 
\begin{equation} \label{eq:coll.sigma}
F = \tilde F - T \Sigma.
\end{equation}
In order to get an explicit expression for $\Sigma$ we introduce replicas.
The replicated free energy reads:
\begin{equation}
F(m) = - \lim_{u \to 0^+} \frac{T}{m} \ln \int \mathcal{D} \psi \,
\left( \tilde Z [\psi] \right)^m,
\end{equation}
from which, evidently, $\tilde F$ can be obtained as
\begin{equation}
\tilde F = \frac{\partial m F(m)}{\partial m} \bigg|_{m=1},
\end{equation}
and hence, according to Eq.~(\ref{eq:coll.sigma})
\begin{equation} \label{eq:coll.sconf}
\Sigma = \frac{1}{T} \frac{\partial F(m)}{\partial m} \bigg |_{m=1}.
\end{equation}
Using Eq.~(\ref{eq:coll.pinning}) and integrating over $\psi$, we get:
\begin{eqnarray} \label{eq:coll.replica}
\nonumber
Z(m) &=& \lim_{u \to 0^+} \int \prod_{a=1}^m \mathcal{D} \phi^a \,
\exp \Bigg ( - \beta \sum_{a=1}^m \mathcal{H}[\phi^a] \\
&& \qquad
+ \, \frac{u}{2m} \sum_{a,b=1}^m \int \drv^3 \mathbf{x} \phi^a (\mathbf{x})
\phi^b (\mathbf{x}) \Bigg).
\end{eqnarray}
The partition function of Eq.~(\ref{eq:coll.replica}) is formally equivalent
to that of a system in a quenched random field analyzed by means
of the replica trick, such as the random field Ising model (RFIM).
The only difference is that in this case the limit
$m \to 1$ must be taken. We can thus use the techniques developed to
deal with such systems~\cite{mezard1}.
The matrix correlation function of the problem of Eq.~(\ref{eq:coll.replica})
obeys the following Dyson equation:
\begin{equation} \label{eq:coll.dyson}
G_{ab}^{-1} (\mathbf{k}) = G_{0}^{-1} (\mathbf{k}) \delta_{ab} +
\Sigma_{ab} (\mathbf{k}) - \frac{u}{m},
\end{equation}
where $G_{0}^{-1} (\mathbf{k})$ is the propagator of the free
unreplicated theory, Eq.~(\ref{eq:coll.gk}), and
$\Sigma_{ab} (\mathbf{k})$ is the self-energy in the replica space.
If we find that, due to the ergodicity-breaking coupling constant $u$,
$\Sigma_{ab} (\mathbf{k})$ has finite off-diagonals elements, we can
conclude that there must be an energy landscape sensitive to that
infinitesimal perturbation, leading to a glassy dynamics. On the other
hand, if the self-energy is diagonal, the dynamics is ergodic and the system
is in the liquid phase.

\subsection{The Self-Consistent Screening Approximation}
In the following we use the self-consistent screening approximation
(SCSA)~\cite{bray,mezard1,wolynes,wolynes1}, which consists in
introducing a $N$-component version of the model with
$\phi = (\phi_1, \ldots, \phi_N)$ and coupling constant $g/N$ and
summing self-consistently
all the diagrams of order $1/N$ (see Fig.~\ref{fig:coll.SCSA}).
This approximation is exact up to order $1/N$.
At the end of calculations we will consider the scalar case, 
$N=1$~\cite{nota_scsa}.
Since the attractive coupling between replicas is symmetric with respect
to the replica index, we assume the following structure of the
matrix correlation function:
\begin{equation} \label{eq:coll.1rsb}
G_{ab} (\mathbf{k}) = \left[ \mathcal{G} (\mathbf{k}) - \mathcal{F}
(\mathbf{k}) \right] \delta_{ab} + \mathcal{F} (\mathbf{k}),
\end{equation}
i.e., with diagonal elements $\mathcal{G} (\mathbf{k})$ and
off-diagonal elements $\mathcal{F} (\mathbf{k})$.
 It can be shown
that for system with quenched disorder (such as the RFIM), this ansatz turns
out to be equivalent to the one-step replica symmetry breaking ansatz (1RSB).
While the diagonal correlator can be
interpreted as the usual (one time) correlation function, i.e.,
\begin{equation}
T \mathcal{G} (\mathbf{k}) = \langle \phi_{\mathbf{k}} \phi_{\mathbf{-k}}
\rangle,
\end{equation}
the off-diagonal term can be interpreted as measuring
the long-time correlations:
\begin{equation}
T \mathcal{F} (\mathbf{k}) = \lim_{t \to
\infty} \langle \phi_{\mathbf{k}} (t) \phi_{\mathbf{-k}} (0) \rangle.
\end{equation}
Hence, $\mathcal{F} (\mathbf{k})$ vanishes in the paramagnetic phase while is
finite in the glassy one.

\begin{figure}
\begin{center}
\psfig{figure=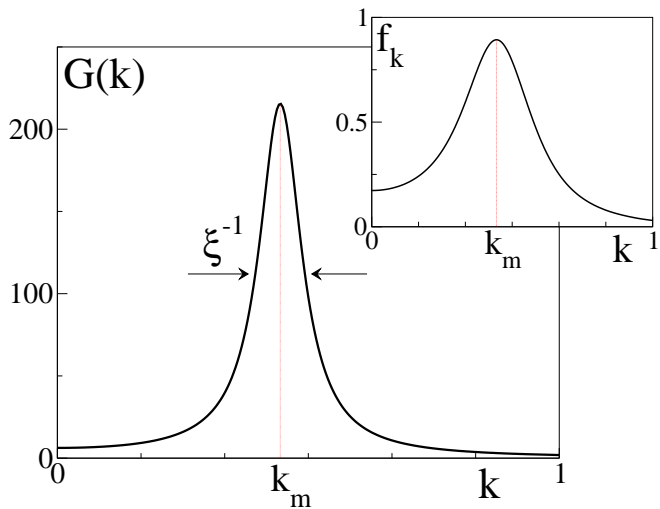,scale=0.31,angle=-90}
\end{center}
\caption{
{\bf Main frame:} Momentum dependence of the correlation function,
$G(\mathbf{k})$, for $4 \pi W = 0.2$ and $\lambda=2$ at $T=T_d$, showing
that it is peaked around the typical modulation wave vector $k_m$ with width
$\xi^{-1}$, given by the inverse of the correlation length, signaling the
fact that microphase structures establish over a finite range $\xi$.
{\bf Inset:} Momentum dependence of the non ergodicity parameter,
$f_{\mathbf{k}}$, for the same values of $W$, $\lambda$ and $T$.
}
\label{fig:coll.gk}
\end{figure}
As shown in Ref.~\cite{wolynes1} and in App.~\ref{app:scsa}, 
within the SCSA the expressions of the correlators can be found by
solving numerically a set of coupled integral equations, 
Eqs.~(\ref{eq:coll.SCSAa})-(\ref{eq:coll.SCSAd2}).
After that, according to Eqs.~(\ref{eq:coll.sc1}) and (\ref{eq:coll.sc2}) we
are able to compute the complexity, $\Sigma$.

\subsection{Results}
We have fixed the range of the repulsive potential to $\lambda = 2$
and we have studied the behaviour of the complexity for different values of the
strength of the repulsion, $W$, and of the temperature $T$.
The model undergoes a glass transition of the same
nature of that found in discontinuous spin 
glasses~\cite{wolynes,wolynes1,letter}.
In Fig.~\ref{fig:coll.sc} the behaviour of the configurational entropy 
is plotted for two different values of $W$ as a function of the temperature.
For a given value of the strength of the repulsion $W$, at high temperature,
$\mathcal{F} (\mathbf{k})$ vanishes, leading to a vanishing
complexity, corresponding to the fact that the system is in the liquid phase.
At $T=T_d(W)$ the configurational entropy jumps discontinuously to a
finite value, signaling the emergence of an
exponentially large number of metastable states. 
At this temperature a glassy dynamics sets in, corresponding to
a non-zero value of the long time correlation function
$\mathcal{F} (\mathbf{k})$.
The complexity
decreases as the temperature is decreased and vanishes at 
$T_K(W)$. At this temperature, called the Kauzmann temperature, an ideal
thermodynamic glass transition to a 1RSB glassy phase takes place.
Note that, in agreement with the results presented in the
previous section, the glass transition disappears for $N \to \infty$, when
the corrections due to the diagrams of order $1/N$ are not taken into
account.

In order to characterize the nature of the glass transition, we
examine the properties of the correlation functions~\cite{letter}.
In the main frame of
Fig.~\ref{fig:coll.gk} the diagonal part of the correlator, ${\cal G}
({\bf k})$, is plotted as a function of the wave vector at the dynamical transition
temperature $T_d$, showing that the correlation function is clearly
peaked around a maximum located in $k_m$, whose value is given in 
Eq.~(\ref{eq:coll.kmax}).
As shown in the next section,
the width of such maximum, $\xi^{-1}$, allows to identify the inverse of
the correlation length. The shape of ${\cal G} ({\bf k})$ indicates that,
although no periodic order occurs ($\langle \phi_{\mathbf{k}_m} \rangle = 0$),
a lamellar structure of wave length $l_m = 2 \pi k_m^{-1}$ over a finite range
$\xi$ is formed. These microphase structures are the analog of the cluster
phase observed in colloidal systems. 
The system forms a mosaic of such modulated
structures, periodically ordered over length scales smaller than the
correlation length $\xi$, and
randomly assembled in a disordered fashion over larger length scales.
The defects and the imperfections in the perfect stripe arrangements 
give rise to the tails in the correlation function.
The lower is the temperature, the more pronounced is the peak of
${\cal G} ({\bf k})$ around the maximum in $k_m$,
signaling the fact that $\xi$ increases as the temperature is decreased.
At $T=T_d$, where $\xi \gtrsim 2 l_m$,
these microphase structures establish over a
length larger than their modulation length and becomes frozen.
The glass transition arises
from the fact that there are many possible configurations to
arrange such modulated structures in a disordered fashion, leading
to the emergence of metastable states.

\begin{figure}
\begin{center}
\includegraphics[scale=0.31,angle=270]{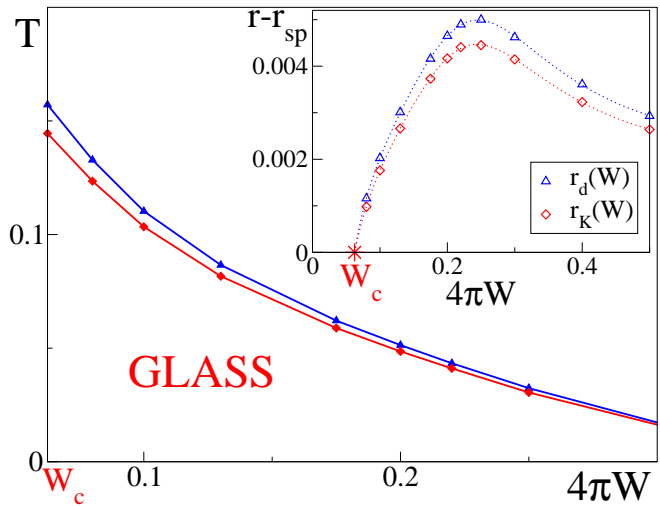}
\end{center}
\caption{(color online) 
{\bf Main frame:} Dynamical (blue curve and triangles) and thermodynamical
(red curve and diamonds) transition temperature to the glassy phase, as a
function of the strength of the repulsive potential, $4 \pi W$, for $\lambda = 
2$. 
{\bf Inset:} Values of the renormalized mass, $r_d$ and $r_K$, corresponding,
respectively, to the dynamical and the thermodynamical glass transition
temperatures, as a function of $4 \pi W$. In the figure the difference
between $r_d(W)$ (red curve and triangles) and $r_K(W)$ (blue curve and 
diamonds) and the minimum value of the renormalized mass, $r_{sp} (W)$, is 
plotted. The figure shows that as $W \to W_c$, both $r_d$ and $r_K$ approach 
$r_{sp} (W)$, signaling that $T_{d,K} (W_c) = 0$. 
}
\label{fig:coll.tk}
\end{figure}
The characteristic wave length, $k_m^{-1}$, dominates also the dynamics
as indicated by the momentum dependence of the non ergodicity
parameter
\begin{equation} \label{eq:coll.nep}
f_{\mathbf{k}} \equiv \lim_{t \to \infty} \frac{\langle
\phi_{\mathbf{k}} (t) \phi_{\mathbf{-k}} (0) \rangle} {\langle
\phi_{\mathbf{k}} (t) \phi_{\mathbf{-k}} (t) \rangle} =
\frac{{\cal F}(\mathbf{k})}{{\cal G}(\mathbf{k})},
\end{equation}
plotted in the inset of
Fig.~\ref{fig:coll.gk} at $T = T_d$.
Basically, $f_{\mathbf{k}}$ represents the height of the
plateau reached by the correlation function of the density fluctuations
of wave vector ${\bf k}$ after infinite time.
From Eq.~(\ref{eq:coll.nep}) one immediately observes that as
${\cal F}(\mathbf{k}) = 0$ ($T>T_d$), also the non ergodicity parameter,
$f_{\mathbf{k}}$, vanishes: the system is in an ergodic phase,
since all the correlation functions relax to zero.
On the other hand, at $T=T_d$ the ergodicity is broken. The presence of
a maximum of $f_{\mathbf{k}}$ located in $|{\bf k}| = k_m$ signals the fact
that structural arrest is more pronounced over length scales of order $l_m$.
The width of the maximum of the non ergodicity parameter, $\Upsilon^{-1}$, can
be tied to the emergence of a new length scales. Indeed,
the height of the plateau decays as $||\mathbf{k}| - k_m|> \Upsilon^{-1}$, 
and $\Upsilon$ represents
the typical length scale over which defects and imperfections in the
periodically modulated pattern are allowed to wander and to diffuse over long
times~\cite{wolynes,wolynes1}.

In Fig.~\ref{fig:coll.tk} the dynamical transition temperature $T_d(W)$ and the
ideal one $T_K(W)$ are plotted as a function of the frustration $4 \pi W$, for
a fixed value of the range of the repulsive potential, $\lambda = 2$.
It is also interesting to study the behaviour of $r_d(W)$ and $r_K(W)$, i.e.,
the values of the renormalized mass $r$ corresponding, respectively, 
to the dynamical and to
the thermodynamical transition temperature. More precisely,
in the inset of Fig.~\ref{fig:coll.tk}, $r_d(W) - r_{sp} (W)$
and $r_K(W) - r_{sp} (W)$ are plotted, where $r_{sp} (W)$ is the minimum value 
of the renormalized mass (corresponding to $T=0$), 
whose expression is given in Eq.~(\ref{eq:coll.rsp}).
The figure clearly shows that
the glassy phase disappears as $W \to W_c$, where we find that $r_{d,K} (W_c)
\to r_{sp} (W_c)$, i.e., $T_d (W_c) = T_K (W_c) = 0$. (Nevertheless, 
we have that $\lim_{W \to W_c^+} T_{d,K} (W) \neq T_{d,K} (W_c) = 0$).
These results can be interpreted if one observes that,
according to Eq.~(\ref{eq:coll.kmax}), the size of the modulated structures
increases as $W$ is decreased and diverges as $W \to W_c$. In fact, for
purely geometrical reasons, the larger is the modulation length, the smaller is
the number of metastable configurations to arrange the microphase
structures in a disordered fashion. As a consequence, the number of metastable
states (i.e., the complexity) vanishes as $W \to W_c$ and the glassy
phase disappears. This is clearly outlined by the inset of
Fig.~\ref{fig:coll.sc}, which shows that the complexity computed at 
the dynamical transition
temperature tends
to zero as $W \to W_c$.

\begin{figure}
\begin{center}
\includegraphics[scale=0.31,angle=270]{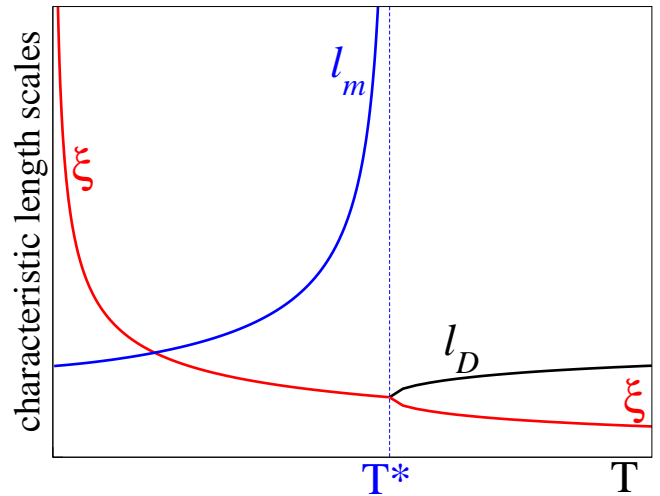}
\end{center}
\caption{(color online) 
Temperature dependence of the characteristic length scales. At high
temperatures, $T > T^{\star}$, the system behaves as a fluid of charges of
linear size $\xi$ (red curve) and screening length $l_D$ (black curve).
At $T^{\star}$ this description breaks down and the competition between
attraction and repulsion produces microphase structures with interstripe
distance $l_m$ (blue curve), established over a finite range $\xi$.
The modulation length, $l_m$, decreases as the temperature is decreased and
at low temperature equals $2 \pi k_m^{-1}$. Conversely, the correlation
length, $\xi$, increases as the temperature is decreased and diverges at $T \to 
0$ at the spinodal line of the homogeneous phase. Glassiness emerges when $\xi
\gtrsim 2 l_m$. On the contrary, the modulated structures progressively fade
as $\xi \lesssim l_m$. 
}
\label{fig:coll.length}
\end{figure}
\section{Modulated structures and characteristic length scales}
\label{sec:length}
In this section we analyze more accurately the nature of the microphase
structures which are formed due to the competition between attraction and
repulsion on different length scales. This investigation provides many insights
on the nature of the cluster phase observed in colloidal suspensions and on the
physical mechanisms inducing the structural arrest in these systems.

Let us consider the case $W > W_c$, where
the correlation function, $G({\bf k})$, 
is peaked around the typical modulation
wave vector, $k_m$. In the following we compute the
correlation function in real space (within the self-consistent Hartree
approximation, or, equivalently, in the spherical limit)
and we show that different length scales emerge in the 
system~\cite{kivelson,nussinov}.
The pair correlator in real space is determined
by computing the Fourier transform of $G({\bf k})$, given in 
Eq.~(\ref{eq:coll.gk}):
\begin{eqnarray} \label{eq:coll.gx}
G(\mathbf{x}) &=& \frac{1}{(2 \pi)^3} \int \drv^3 \mathbf{k} \, G(\mathbf{k})\,
e^{i \mathbf{k} \cdot \mathbf{x}} \\
\nonumber & = &
\frac{T}{2 \pi^2 |\mathbf{x}|} \int \drv k \,
\frac{k (\lambda^{-2} + k^2) \sin (kx)}{(k^2 + \zeta^2)(k^2 + \kappa^2)},
\end{eqnarray}
where
\begin{equation} \label{eq:coll.poles}
\zeta^2, \kappa^2 = \frac{ (r+\lambda^{-2}) \pm \sqrt{(r - \lambda^{-2})^2 -
16 \pi W}}{2}.
\end{equation}
The above equation marks a crossover temperature $T^{\star}$, corresponding
to the temperature at which the renormalized mass equals:
\begin{equation}
r^{\star} = \lambda^{-2} + 2 \sqrt{4 \pi W}.
\end{equation}

\subsection{High temperature: $r > r^{\star}$}
For high enough temperatures, $r > r^{\star}$ (i.e., $T > T^{\star}$), 
the argument of the square
root of the right hand side of Eq.~(\ref{eq:coll.poles}) is positive.
Consequently, $\zeta^2$ and $\kappa^2$ assume real values and
the integral, Eq.~(\ref{eq:coll.gx}), can be easily evaluated by applying the
residue theorem to the
poles lying on the imaginary axis at $k=\pm i\zeta, \pm i \kappa$, leading to
\begin{equation} \label{eq:coll.corr}
G(\mathbf{x}) = \frac{T \left[
(\zeta^2 - \lambda^{-2}) \, e^{- \zeta |\mathbf{x}|}
+ (\lambda^{-2} - \kappa^2) \, e^{- \kappa |\mathbf{x}|} \right]}
{4 \pi |\mathbf{x}| (\zeta^2 - \kappa^2)}.
\end{equation}
The above expression clearly indicates the emergence of two characteristic
length scales, $\xi = |\Re\{\zeta\}|^{-1}$ and $l_D = |\Re\{\kappa\}|^{-1}$.
The former plays the role of the correlation length of the canonical
short-range ferromagnet and at high temperature behaves as:
\begin{equation}
\xi \sim r^{-1/2}.
\end{equation}
The other length $l_D$, instead, is a renormalized screening length,
and for high temperature tends to
\begin{equation}
l_D \sim \lambda,
\end{equation}
the range of the repulsive potential. Therefore, the system at high
temperatures, $T > T^{\star}$,
behaves as a fluid of charges of linear size $\xi$ and
screening length $l_D$.
The temperature dependence of $\xi$ and $l_D$ is shown in
Fig.~\ref{fig:coll.length}.

\subsection{Low temperature: $r < r^{\star}$}
\begin{figure}
\begin{center}
\includegraphics[scale=0.3175,angle=270]{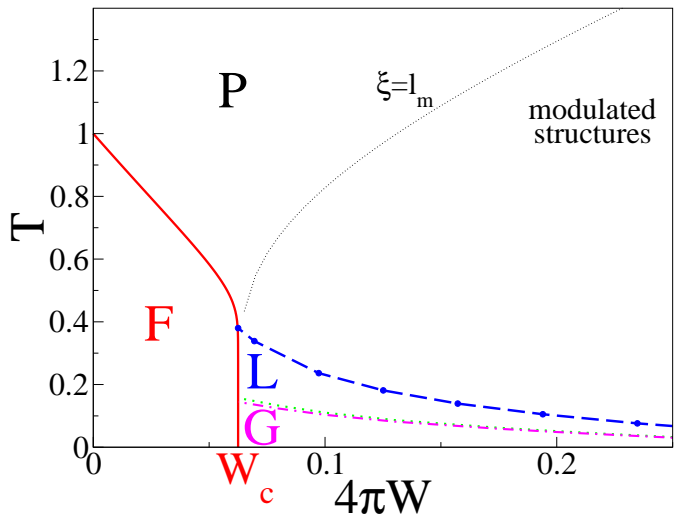}
\end{center}
\caption{
System phase diagram in the frustration ($4 \pi W$)-temperature ($T$) plane,
for a fixed value of the screening length of the repulsive potential, $\lambda
= 2$, and for $g=1$ and $r_0 = -1$. The temperature is scaled with respect to
the critical temperature of the unfrustrated system, $T_c (W = 0) = - 2 \pi^2
r_0/3 g \Lambda$. 
For $W<W_c$ a second order
phase transition with the usual Hartree critical exponents from a paramagnetic
phase (P) to a ferromagnetic (F) one is found, by lowering the temperature
below $T_c (W, \lambda)$ (red continuous curve). In the terminology of 
colloidal system this corresponds to a phase separation between a colloid rich
and a colloid poor phase. For $W>W_c$ the phase separation is prevented and,
instead, a first order transition to a lamellar phase (L), characterized by a
periodic variation of the density, is found at $T_L (W, \lambda)$ (blue dashed
line). Such a first order transition can be avoided and, in this case, at lower
temperature a transition to a glassy phase (G) is found, characterized by a
dynamical, $T_d (W, \lambda)$ (green dotted curve), and by a thermodynamical,
$T_K (W, \lambda)$ (magenta dashed-dotted curve) transition. The black dotted
line represent a crossover temperature, corresponding to the temperature at
which the correlation and the modulation lengths equal, $\xi \simeq l_m$, and
the system establishes microphase structures, which are the analog of the
clusters observed in colloidal systems. 
}
\label{fig:coll.diagram}
\end{figure}

Conversely, for $r < r^{\star}$ (i.e., $T<T^{\star}$) 
the argument of the square root of
Eq.~(\ref{eq:coll.poles}) becomes negative.
Consequently, $\zeta^2$ and $\kappa^2$ assume complex values.
The analytic continuation of Eq.~(\ref{eq:coll.corr}) to the low temperature
region reads:
\begin{eqnarray} \label{eq:coll.modul}
\nonumber
G(\mathbf{x}) &=& \frac{T}{16 \pi |\mathbf{x}|\kappa_1 \kappa_2} \,
e^{-\kappa_1 |\mathbf{x}|} 
\Big[ \left( 2\kappa_1\kappa_2 - 
\lambda^{-2} \right) \cos (\kappa_2 |\mathbf{x}|) \\
&& \qquad \,\, +\, 
\left(\kappa_2^2 - \kappa_1^2 -
\lambda^{-2} \right) \sin (\kappa_2 |\mathbf{x}|) \Big],
\end{eqnarray}
where $\kappa = \kappa_1 + i \kappa_2$. 
This expression implies that, although no
periodic order occurs ($\langle \phi_{\mathbf{k}_m} \rangle = 0$),
a lamellar structure of wave length $l_m$ over a
finite range $\xi$ is formed. 
The modulation length is given by:
\begin{equation}
l_m =  2\pi|\kappa_2|^{-1} =
4\pi \left[ 2 \sqrt{r \lambda^{-2} + 4 \pi W}
- r - \lambda^{-2} \right]^{-1/2},
\end{equation}
whereas, according to Eq.~(\ref{eq:coll.modul}),
the correlation length reads:
\begin{equation}
\xi = |\kappa_1|^{-1} = 2 \left[ \left(r+\lambda^{-2} \right) + 2 \sqrt{r
\lambda^{-2} + 4 \pi W} \right]^{-1/2},
\end{equation}
and gives the range over which the microphase structures are formed.
The temperature dependence of the characteristic length scales present in the
system is shown in Fig.~\ref{fig:coll.length}.
The correlation length increases as the temperature is decreased and
diverges as $T$ goes to zero, where the spinodal line of the paramagnetic
phase is located. 
At low temperatures we have $l_m = 2 \pi k_m^{-1}$, where $k_m$ is the
characteristic wave vector defined in Eq.~(\ref{eq:coll.kmax}). Conversely, as
$T \to T^{\star -}$, the modulation length
$l_m$ diverges as $(r^{\star} - r)^{-1/2}$.
Interestingly enough, it turns out that the glass transition emerges at
temperatures at which $\xi \ge 2 l_m$, as conjectured in Ref.~\cite{wolynes1}:
the microphase structures form over a
length larger than their modulation length and become frozen. The
glass transition arises from the fact that there are many possible
configurations to arrange such modulated structures in a disordered
fashion, leading to a great number of metastable states.
As the temperature is increased, the cluster phase continuously fades
approximately at the temperature at which $\xi \lesssim l_m$: the microphase
structures are ordered over a length scale smaller than their own modulation
length and the system becomes homogeneous. Such a crossover temperature at
which approximately the modulated structures fades and the system is
homogeneous corresponds to the black dotted line of
Fig.~\ref{fig:coll.diagram}.

\section{Phase Diagram and Concluding remarks} \label{sec:phdi}
The results found in the previous sections are summarized in
Fig.~\ref{fig:coll.diagram}, where the complete phase diagram of the system is
presented, showing the relative position of the different phases.
The model, 
albeit schematically,
retains the essential
physics of charged colloids in polymeric solutions, where the competition
between attraction and repulsion on different length scales plays a prominent
role.
The results found
provide many insights on the equilibrium phase diagram of colloidal
suspensions, as well as on the physical mechanisms
responsible for the formation
of modulated structures and for the structural arrest.

We point out that in our model, for $W>W_c$, the competition between 
attraction and repulsion has the effect to produce microphase structures,
which are the analog of the cluster phase observed in colloidal system,
characterized by a peak in the structure factor around a wave vector
corresponding to the inverse of the typical size of the clusters. 
The width 
of the peak is related to the correlation length, which represents the range
over which such modulates structures are locally ordered.
In our model the modulated structures are the precursors of
an equilibrium lamellar phase occurring at lower temperature.
Therefore, according to this analogy, our
results suggest that the cluster phase observed
in colloidal suspensions might be the signature
of the presence in the system phase diagram of an equilibrium lamellar phase,
which is very often kinetically avoided on the experimental time scales.
Such an underlying equilibrium ordered phase may strongly
affect the dynamics of the system in the low temperature region,
close to the colloidal gelation.
The presence of a stripe/columnar/lamellar phase in the phase diagram
and the emergence of microphase structures 
has been clearly shown in recent numerical simulations of model
systems of particles interacting via the DLVO potential both in 
two~\cite{char_reich,reatto} and in three dimensions~\cite{tubi},
where the presence of first order transitions from a cluster phase 
to periodically ordered phases has been clearly detected.

The nature of the glass transition found in our model gives many
insights on the nature of the mechanisms inducing the phenomenon of colloidal
gelation. We suggest that the
structural arrest in colloidal suspensions might be due to the formation of
microphase structures, which are ordered on small length
scales, and are assembled together in a disordered fashion on larger
length scales, thereby inducing a complex free energy landscape and,
consequently, a complex and slow dynamics.

Notice that in our model there is no geometric frustration due to the hard-core
excluded volume, which is totally absent in the Hamiltonian. 
The frustration lies only in the competition between
attractive and repulsive interactions on different length scales. Therefore,
our results seem to indicate that the physical mechanism inducing the 
structural arrest 
in charged colloids is totally different from that responsible
for the jamming and the glass transition in molecular liquids at high 
volume fraction.

In real experiments the addition of salt increases the number of ions in
the solution, thereby decreasing the strength of the repulsive shoulder
of the potential. Our results suggest that there is a critical value
of the ions concentration (corresponding to $W_c$) above which no modulated
structures are formed and
the system undergoes, instead, a phase separation between a colloid-rich and
a colloid-poor region.

All these prediction could be experimentally and numerically checked.

~

We   
warmly thank A. de Candia, E. Del Gado,
A. Fierro, G. Gonnella and N. Sator. M.T. is grateful G.
Tarjus for many useful discussions and remarks.
Work supported by  EU Network Numbers HPRN-CT-2002-00307 and 
MRTN-CT-2003-504712,
MIUR-PRIN 2004, MIUR-FIRB 2001, CRdC-AMRA, INFM-PCI.

\appendix

\section{Critical exponents of the ferromagnetic transition in
the Hartree approximation} \label{app:hart}
In this appendix we compute
the critical exponents of the second order ferromagnetic phase transition
found within the Self-Consistent Hartree approximation for $W < W_c$.

As shown in Sec.~\ref{sec:hart}, for $W < W_c$,
the correlation function, $G({\bf k})$, has a maximum in $|{\bf k}| = 0$, where
it equals $T/(r + 4 \pi W \lambda^2)$. Therefore, as $r \to r_c = - 4 \pi W
\lambda^2$, the susceptibility, $\chi = G(|{\bf k}| = 0)/T$ diverges,
corresponding to a second order critical point.
Since the divergence occurs around $|{\bf k}| = 0$, in order to determine the
critical behaviour, we can expand the Hamiltonian of Eq.~(\ref{eq:coll.model})
up to the lowest orders in $k^2$, as done in Eq.~(\ref{eq:coll.exp}), 
neglecting the other irrelevant terms. According
to Eq.~(\ref{eq:coll.gk}),
the susceptibility reads:
\begin{equation}
\chi^{-1} (r) = r + 4 \pi W \lambda^2 = r - r_c \equiv \tau.
\end{equation}
Replacing $r$ and $r_c$ with their expressions obtained via the self-consistent
relation, Eq.~(\ref{eq:coll.scha}), and expanding the correlators up to
$O(k^2)$, we get:
\begin{eqnarray} \label{eq:ah.eq}
\nonumber
\tau &=& \frac{3 g}{2 \pi^2} \int_0^{\Lambda} \drv k \, k^2 \bigg[
\frac{T}{r + 4 \pi W \lambda^2 + (1 - 4 \pi W \lambda^4) k^2} \\
&& \qquad \,\, - \,
\frac{T_c}{(1 - 4 \pi W \lambda^4) k^2} \bigg].
\end{eqnarray}
Now, we define the function $I(\tau)$ as
\begin{equation}
I (\tau) = \int_0^{\Lambda} \drv k \, \frac{1}{\tau + c k^2}.
\end{equation}
Recalling that $\tau = r - r_c = r  + 4 \pi W \lambda^2 = 1 - W/W_c$ and
that $c = 1 - 4 \pi W \lambda^4$ (note that $c=0$ for $W=W_c$),
in terms of the function $I(\tau)$ Eq.~(\ref{eq:ah.eq}) can be written as:
\begin{equation} \label{eq:ah.exp1}
\tau = \frac{3 g (T - T_c)}{2 \pi^2} \int_0^{\Lambda} \drv k \,
\frac{c k^2}{\tau + c k^2} \,
- \, \frac{3 g T_c}{2 \pi^2 c} \, \tau I(\tau).
\end{equation}
The first integral appearing in the previous equation approaches a constant as
$\tau \to 0$, whereas the integral $I(\tau)$ is divergent as $\tau \to 0$ as
\begin{equation}
I(\tau) \sim \tau^{-1/2}.
\end{equation}
To the leading order in $\tau$, the left hand side of Eq.~(\ref{eq:ah.exp1})
can be neglected, leading to
\begin{equation}
T - T_c \sim \tau^{1/2},
\end{equation}
and, thus:
\begin{equation}
\chi^{-1} = \tau^{-1} \sim (T - T_c)^{-2}.
\end{equation}
This allows to determine the exponent $\gamma$:
\begin{equation}
\gamma = 2.
\end{equation}
Within the self-consistent Hartree approximation, there are no ${\bf
k}$-dependent corrections to the mean field propagator, all the corrections
being included in the renormalized mass, $r$. Therefore,
\begin{equation}
\chi ({\bf k}) = \left[ \tau + (1 - 4 \pi W \lambda^4) k^2 \right]^{-1}
= \chi \left[1 + (k \xi)^2\right]^{-1},
\end{equation}
where
\begin{equation}
\xi^2 = (1 - 4 \pi W \lambda^4)/\tau \sim (T - T_c)^{-2}.
\end{equation}
This immediately leads to
\begin{equation}
\nu = \frac{\gamma}{2} = 1.
\end{equation}
Similarly, it is possible to verify that
also for $W>W_c$, at the spinodal line of the paramagnetic phase, located at
$T=0$, the susceptibility, $\chi (k_m)$, and the correlation length, $\xi$,
diverge with the usual Hartree critical exponents.

\section{Fluctuatuion-induced first order transition to the lamellar phase}
\label{app:fifot}
In this appendix we determine the first order transition temperature 
$T_L (W, \lambda)$ from the homogeneous phase to the lamellar phase. 
We follow the approach taken in~\cite{tarjus1} for the 
Coulomb case and we generalize it to the case of a finite screening
length of the repulsive potential. Let us
introduce a spatially varying external field, $h_{\bf k}$, linearly coupled
to the order parameter field $\phi_{\bf k}$. As a result, $\phi_{\bf k}$ is
now the sum of an average component,
$m_{\bf k} =  \langle \phi_{\bf k} \rangle$,
and a fluctuation around it $\xi_{\bf k}= \phi_{\bf k} -m_{\bf k}$.
The resulting equation of state for the average components 
reads~\cite{tarjus1}:
\begin{eqnarray} \label{eq:coll.a1}
&& h_{\bf k} = \left(r_0+k^2+\frac{4 \pi W}{\lambda^{-2}+ k^2}\right)
m_{\bf k} \\
\nonumber
&& \,\, 
+g\int \frac{\drv^3{\bf k}_1}{(2\pi)^3} \, \frac{\drv^3{\bf k}_2}{(2\pi)^3} \,
\left[m_{{\bf k}_1}
m_{{\bf k}_2}+3 G({\bf k}_1,{\bf k}_2)\right]m_{{\bf k}-{\bf k}_1-{\bf k}_2},
\end{eqnarray}
where the connected correlation function
$G({\bf k},{\bf k}^{\prime})=\langle \xi_{{\bf k}} \, \xi_{{\bf k}^{\prime}}
\rangle$, is obtained self-consistently by solving
\begin{eqnarray}\label{eq:coll.a2}
\!\!\!\! && \!\!\!\!
T G^{-1}({\bf k},{\bf k}^{\prime})=\left(r_0+k^2+\frac{4 \pi W}{\lambda^{-2} +
k^2} \right)\delta({\bf k}+{\bf k}^{\prime})\\
\nonumber
&& \,\,\,\,\,\,\,\, 
+ \, 3 g \int \frac{\drv^3{\bf q}}{(2\pi)^3}\Big[m_{ \bf q} m_{{\bf k}
+{\bf k}^{\prime}-{\bf q}}
+ G({\bf q},{\bf k}+{\bf k}^{\prime}-{\bf q})\Big],
\end{eqnarray}
together with the unitariety condition:
\begin{equation}
\int \frac{\drv^3{\bf q}}{(2\pi)^3} \, G^{-1}({\bf k},{\bf q}) \,
G({\bf q},{\bf k}^{\prime}) = \delta ( {\bf k} - {\bf k}^{\prime}).
\end{equation}
Note that in the  paramagnetic phase, when all $h_{\bf k}$, and
$m_{\bf k}$ are equal to zero, Eqs.~(\ref{eq:coll.a1}) and
(\ref{eq:coll.a2}) reduce to Eq.~(\ref{eq:coll.scha}) with
$G({\bf{k}},{\bf{k}^{\prime}})=G({\bf{k}}) \,
\delta ({\bf k}+{\bf k}^{\prime})$.

In the lamellar phase, characterized by the periodic order,
instead, we consider:
\begin{equation} \label{eq:coll.a3}
h_{ \bf k}=h\big(\delta({\bf k}-{\bf k}_m)+\delta({\bf k}+ {\bf k}_m)\big),
\end{equation}
and
\begin{equation}\label{eq:coll.a4}
m_{\bf k}=m\big(\delta({\bf k}-{\bf k}_m)+\delta({\bf k} +{\bf k}_m)\big),
\end{equation}
where ${\bf k}_m$ is given in Eq.~(\ref{eq:coll.kmax}).
As shown by Brazovskii for a related model~\cite{brazov},
in this region, the fluctuations of wave-vector ${\bf k}$ with
$|{\bf k}|=k_m$ are dominant, and the
effect of the off-diagonal terms with ${\bf k} \neq {\bf k}^{\prime}$ can be
neglected in the correlation function. As a result,
\begin{equation}
T G^{-1} (\mathbf{k}) = r + k^2 + \frac{4 \pi W}{\lambda^{-2} + k^2}.
\end{equation}
This expression is formally equivalent to that of Eq.~(\ref{eq:coll.scha}). 
However, in the case of a phase characterized by
Eqs.~(\ref{eq:coll.a3}) and (\ref{eq:coll.a4})
the renormalized mass, $r$, is given by:
\begin{eqnarray} \label{eq:coll.a5}
r &= & r_0+3g\int\frac{\drv^3{\bf k}}{(2\pi)^3} \left (
G (\mathbf{k}) + |m_{\bf k}|^2 \right) \\
\nonumber
& = & r_0+3g\int\frac{\drv^3{\bf
k}}{(2\pi)^3}\,\frac{T}{r+k^2+\frac{4 \pi W}{\lambda^{-2} + k^2}}+6g|m|^2.
\end{eqnarray}
By introducing Eqs.~(\ref{eq:coll.a3})-(\ref{eq:coll.a5}) in
Eq.~(\ref{eq:coll.a1})
and recalling Eq.~(\ref{eq:coll.kmax}), one obtains the following equation
of state:
\begin{eqnarray}\label{eq:coll.a6}
\nonumber
h&=&\bigg(r_0+2\sqrt{4 \pi W} - \lambda^{-2} \\
\nonumber
&& \,\,\, + \, 3g \int\frac{\drv^3{\bf
k}}{(2\pi)^3}\,\frac{T}{r+k^2+\frac{4 \pi W}{\lambda^{-2} + k^2}}+
3g|m|^2\bigg)m\\
&=&\left(r+ 2\sqrt{4 \pi W} - \lambda^{-2} -3g|m|^2\right)m.
\end{eqnarray}
When we do not consider fluctuations ($r \to r_0$),
the last equation gives the mean field result.

Below some temperature, there is a
coexistence of the paramagnetic phase and the lamellar phase. In zero
field ($h=0$), the  former  is characterized  by $m=0$
and the  latter  by $m\neq 0$,  where $m$  is solution of
Eq.~(\ref{eq:coll.a6}), i.e., $(r+  2\sqrt{4 \pi W} - \lambda^{-2} -3g|m|^2)=0$.
The  transition  point,
which is then associated with a first order transition, is obtained as
the temperature at which the free energies of the two phases are equal.
It is convenient to calculate
directly the free energy difference $\Delta F(T)$  between the lamellar
($m\neq 0$) and the paramagnetic ($m= 0$) phase at a
given temperature $T$~\cite{tarjus1}:
\begin{equation}\label{eq:coll.a7}
\Delta F=\int_{0}^{m}  \drv m^{\prime} \frac{\partial F}{\partial m^{\prime}}=
2\int_{0}^{m} \drv m^{\prime} h(m^{\prime})
\end{equation}
where $h(m^{\prime})$  is given  by  Eq.~(\ref{eq:coll.a6}).
Changing the integration variable from $m^{\prime}$ to
$r^{\prime} (m^{\prime})$ solution of Eq.~(\ref{eq:coll.a5}),
after some algebra we obtain:
\begin{eqnarray} \label{eq:coll.first}
\nonumber
g \Delta F &=& \int_{r(m=0)}^{r(m)} \drv r^{\prime} \Bigg(
\frac{r^{\prime} + r_0}{2} + 2\sqrt{4 \pi W} - \lambda^{-2} \\
& & \,\,
 + \, \frac{3gT}{4 \pi^2}
\int \drv k \, \frac{k^2}{r^{\prime} + k^2 + \frac{4 \pi W}{\lambda^{-2}+k^2}}
\Bigg)\\
\nonumber
& & \,\,
\times \, \left( \frac{1}{6} + \frac{gT}{4 \pi^2} \int \drv k \, \frac{k^2}
{\left( r^{\prime} + k^2 + \frac{4 \pi W}{\lambda^{-2} + k^2} \right)^2}
\right),
\end{eqnarray}
where $r(m=0)$ is the solution of Eq.~(\ref{eq:coll.scha}) whereas $r(m)$ and
$m$ are solutions of the following coupled equations:
\begin{equation} \label{eq:fifoptu}
\left \{ \begin{array}{l}
r+  2\sqrt{4 \pi W} - \lambda^{-2} -3g|m|^2 = 0 \\
r= r_0+3g\int\frac{\drv^3{\bf
k}}{(2\pi)^3}\,\frac{T}{r+k^2+\frac{4 \pi W}{\lambda^{-2} + k^2}}+6g|m|^2.
\end{array}
\right.
\end{equation}
By solving  Eqs.~(\ref{eq:coll.first}) and (\ref{eq:fifoptu}) 
numerically for several  values  of $W$ and
$T$ (and for $g=1$, $r_0 = -1$ and $\lambda=2$),
we have found  that, for any given value of $W$,
the sign of $\Delta  F$ changes  at a
finite value of $T=T_L$, that  marks the first order
transition between paramagnetic and lamellar phases. The results 
are shown in Figs.~\ref{fig:coll.hart} and \ref{fig:coll.diagram}.
We also solved Eqs.~(\ref{eq:coll.a7})-(\ref{eq:fifoptu}) 
for several values of $W$ and $\lambda$
keeping the temperature $T$ fixed. The transition line in the frustration
($W$)-range ($\lambda$) plane is shown in the inset of
Fig.~\ref{fig:coll.hart}.

\section{Self-consistent solution for the dynamics in the spherical limit}
\label{app:dynamics}
In this section we discuss the self-consistent solution 
of the dynamics of the model in the spherical approximation. More
precisely, our aim is to solve self-consistently 
Eqs.~(\ref{eq:coll.solSK}) and (\ref{eq:Qt}).
In order to do so
we remember that:
\begin{equation} \label{eq:coll.selfconsistent}
\dot Q (t) = r + g \int \! \frac{\drv^3
\mathbf{k}}{(2 \pi)^3} \, G (\mathbf{k}, t).
\end{equation}
From Eq.~(\ref{eq:coll.solSK}),
we note that $G (\mathbf{k}, t)$ has a maximum at a wave vector $k_{
max}(t)$ given by:
\begin{equation} \label{eq:coll.max}
Q(t) = - 2 k_{max}^2(t) t - \frac{4\pi W \lambda^{-2} t}{(\lambda^{-2}
+ k_{max}^2(t))^2},
\end{equation}
which gives the expression of $Q(t)$ in terms of the peak position
$k_{max}(t)$.
Inserting Eq.~(\ref{eq:coll.max}) into Eq.~(\ref{eq:coll.solSK}) we obtain that
$G (\mathbf{k}, t) = G (\mathbf{k}, 0) \, e^{-M t f(k,t)}$, where
\begin{eqnarray}
f(k,t) & = & \frac{4\pi W k^2}{\lambda^{-2} + k^2} + k^4 - 
2 k^2 k_{max}^2(t)\\
\nonumber
&& \,\,\,\, - \, 
\frac{4 \pi W \lambda^{-2} k^2}{(\lambda^{-2} + k_{max}^2(t))^2}.
\end{eqnarray}
Obviously, $f(k,t)$ has a minimum for $k = k_{max}(t)$.

Now we can evaluate the integral in Eq.~(\ref{eq:coll.selfconsistent}) by means
of the saddle point method, provided that $Mt$ is large. This yields:
\begin{eqnarray}
\dot Q(t) &=& r + B \, \frac{k_{max}^2(t)}{\sqrt{M t f^{\prime \prime}
(k_{max},t)}} \\
\nonumber
&& \times \, \exp \left[ M t \left( 1 -\frac{4 \pi W}
{(\lambda^{-2} + k_{max}^2(t))^2} \right)
k_{max}^4(t)\right],
\end{eqnarray}
where $B = 2 g \left( \pi /2 \right)^{3/2} G (k_{max}, 0)$.
Now, we can relate $\dot Q(t)$ to $k_{max}(t)$ through
Eq.~(\ref{eq:coll.max}) and,
inserting the result in the previous equation, we obtain 
Eq.~(\ref{eq:coll.equation}).
The function $P$ equals:
\begin{eqnarray} \label{eq:P}
\nonumber
P &=& \frac{1}{B} \bigg[-r - 2 k_{max}^2(t)  - \frac{4\pi W
\lambda^{-2}}{(\lambda^{-2} + k_{max}^2(t))^2}\\
&& \qquad - \, 4 t k_{max}(t) F(k_{max}(t))
\frac{\drv k_{max}(t)}{\drv t} \bigg],
\end{eqnarray}
where $F (k_{max}(t))$ is defined as:
\begin{equation}
F \left(k_{max}(t)\right) = \left( 1 - \frac{4 \pi W
\lambda^{-2}}
{(\lambda^{-2} + k_{max}^2(t))^3}\right).
\end{equation}
Inserting the right hand side of Eq.~(\ref{eq:coll.equation}) into the
expression of the time dependent structure factor we obtain:
\begin{equation} \label{eq:coll.Gkt}
G (\mathbf{k}, t) = G (\mathbf{k}, 0) \, 
\left[ \frac{P \sqrt{M t \, F\left(k_{max}(t)\right)}}{k_{max}(t)} \right] 
e^{-Mt \Delta f(k,t)},
\end{equation}
where $\Delta f (k,t) = f(k,t) - f(k_{max},t)$.
The position of the peak
is given by the solution of Eq.~(\ref{eq:coll.equation}). Taking the logarithm
of both sides of Eq.~(\ref{eq:coll.equation}) and dividing by $Mt$ we obtain
an asymptotic expression
for $k_{max}(t)$ which is readily analyzed in the limit $t \to \infty$.
\begin{eqnarray} \label{eq:coll.asympt}
\!\!\!\! && \!\!\!\! \left( 1 -\frac{4 \pi W}
{(\lambda^{-2} + k_{max}^2(t))^2} \right) k_{max}^4(t) \\
\nonumber
&& \qquad \, = \, \frac{ \ln \left[-r - 2 k_{max}^2(t)  - \frac{4\pi W
\lambda^{-2}}{(\lambda^{-2} + k_{max}^2(t))^2} \right]}{Mt}\\
\nonumber
&& \qquad \qquad \,\,\,\,\, 
+ \, \frac{1}{2} \, \frac{\ln \left[F(k_{max} (t))\right]}{Mt} -
\frac{\ln \left[ k_{max} (t)\right]}{Mt},
\end{eqnarray}
where we have neglected the terms which vanish in the asymptotic limit, $t \to
\infty$. As explained in Sec.~\ref{sec:dyn}, 
depending on the values of $W$ and $\lambda$,
different solutions of the previous equation are found. In particular, one
needs again to distinguish between the cases $W<W_c$ and $W>W_c$.

\section{Self-consistent screening approximation}\label{app:scsa}
Here we discuss how to compute the complexity within the 
SCSA. This section is unessential, since all the calculations have been
already extensively reported in the literature (see Ref.~\cite{wolynes1}). 
Nevertheless, we believe that this appendix might be useful 
for the sake of completeness.

Following Ref.~\cite{wolynes1}, 
we first
summarize a few properties of the matrices in the replica space with structure
similar to Eq.~(\ref{eq:coll.1rsb}). Introducing a matrix ${\bf E}$ such that
$E_{ab} = 1$ and the unitary matrix ${\bf 1} = \delta_{ab}$, it is easy to see
that the product of any two $m \times m$ matrices with structure:
\begin{equation}
{\bf A} = a_1 {\bf 1} + a_2 {\bf E}
\end{equation}
is given by:
\begin{equation} \label{eq:coll.p1}
{\bf A} \cdot {\bf B} = \left( a_1 b_1 \right) {\bf 1} + \left( a_1 b_2 + a_2
b_1 + m a_2 b_2 \right) {\bf E}.
\end{equation}
This relation leads to
\begin{equation} \label{eq:coll.p2}
{\bf A}^{-1} = \frac{1}{a_1} \, {\bf 1} - \, \frac{a_2}{a_1 \left( a_1 + m a_2
\right)}\, {\bf E}
\end{equation}
for the inverse of a matrix ${\bf A}$. These properties will be used in the
following.

\begin{figure}[t!]
\begin{center}
\includegraphics[scale=0.27]{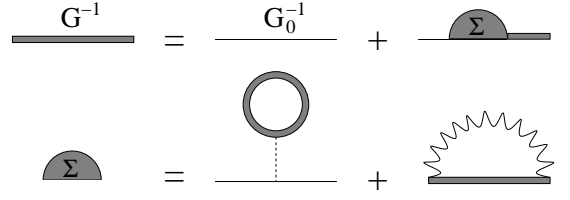}
\end{center}
\caption{Diagrams of the Self-Consistent Screening Approximation
\cite{bray,mezard,wolynes,wolynes1}.}
\label{fig:coll.SCSA}
\end{figure}

In the SCSA, the self-energy, $\Sigma_{ab}$ is given by (see
Fig.~\ref{fig:coll.SCSA}):
\begin{equation} \label{eq:coll.SCSA1}
\Sigma_{ab} (\mathbf{k}) = \frac{2}{N} \int \frac{\drv^3 \mathbf{q}}
{(2 \pi)^3} \,
G_{ab} (\mathbf{k}+\mathbf{q}) D_{ab} (\mathbf{q}),
\end{equation}
where
\begin{equation}
D_{ab} (\mathbf{q}) = \left[ \left( gT \right)^{-1} + \Pi_{ab} (\mathbf{q})
\right]^{-1}
\end{equation}
is determined self-consistently by the polarization function
\begin{equation}
\Pi_{ab} (\mathbf{q}) = \int \frac{\drv^3 \mathbf{p}} {(2 \pi)^3} \,
G_{ab} (\mathbf{q} + {\bf p}) \, G_{ba} (\mathbf{p}).
\end{equation}
The ansatz Eq.~(\ref{eq:coll.1rsb}) for the correlation functions implies an
analogous structure for $\Sigma_{ab} (\mathbf{k})$ and $\Pi_{ab} (\mathbf{k})$.
Inserting this ansatz into $\Pi_{ab} (\mathbf{q})$ gives:
\begin{equation}
\Pi (\mathbf{q}) = \left [ \Pi_{\cal G} (\mathbf{q}) - \Pi_{\cal F}
(\mathbf{q}) \right] {\bf 1} +  \Pi_{\cal F} (\mathbf{q}) {\bf E}.
\end{equation}
where the diagonal and off-diagonal elements of the polarization function are:
\begin{eqnarray} \label{eq:coll.SCSAa}
\Pi_{\cal G} (\mathbf{q}) &=& \int \frac{\drv^3 \mathbf{p}} {(2 \pi)^3} \,
{\cal G} (\mathbf{q} + {\bf p}) \, {\cal G} (\mathbf{p}),\\
\nonumber
\Pi_{\cal F} (\mathbf{q}) &=& \int \frac{\drv^3 \mathbf{p}} {(2 \pi)^3} \,
{\cal F} (\mathbf{q} + {\bf p}) \, {\cal F} (\mathbf{p}).
\end{eqnarray}
Using the property expressed in
Eq.~(\ref{eq:coll.p1}), it is now straightforward to
determine $D_{ab} ({\bf q})$ which, in the limit $m \to 1$, is given by:
\begin{equation}
D_{\cal G} 
(\mathbf{q}) = \left [ {\cal D}_{\cal G} (\mathbf{q}) - {\cal D}_{\cal F}
(\mathbf{q}) \right] {\bf 1} +  {\cal D}_{\cal F} (\mathbf{q}) {\bf E},
\end{equation}
where
\begin{eqnarray} \label{eq:coll.SCSAb}
{\cal D}_{\mathcal{G}} (\mathbf{q}) &=& \left[ \left( gT \right)^{-1} + \Pi_{\cal G}
(\mathbf{q}) \right]^{-1},\\
\nonumber
{\cal D}_{\mathcal{F}} (\mathbf{q}) &=& -
\, \frac{\Pi_{\mathcal{F}} (\mathbf{q})
{\cal D}_{\cal G}^{2} (\mathbf{q})}{1 - \Pi_{\mathcal{F}} (\mathbf{q})
{\cal D}_{\cal G} (\mathbf{q})}.
\end{eqnarray}
Analogously, inserting the above equation into Eq.~(\ref{eq:coll.SCSA1}), we
get for the self-energies
\begin{equation}
\Sigma (\mathbf{q}) = \left [ \Sigma_{\cal G} (\mathbf{q}) - \Sigma_{\cal F}
(\mathbf{q}) \right] {\bf 1} +  \Sigma_{\cal F} (\mathbf{q}) {\bf E},
\end{equation}
where
\begin{eqnarray} \label{eq:coll.SCSAc}
\Sigma_{\cal G} (\mathbf{q}) &=& \frac{2}{N}
\int \frac{\drv^3 \mathbf{p}} {(2 \pi)^3} \, {\cal D}_{\cal G} (\mathbf{p}) \,
{\cal G} (\mathbf{q} + {\bf p}),\\
\Sigma_{\cal F} (\mathbf{q}) &=& \frac{2}{N} \int \frac{\drv^3 \mathbf{p}} {(2
\pi)^3} \, {\cal D}_{\cal F} (\mathbf{p}) \,
{\cal F} (\mathbf{q} + {\bf p}).
\end{eqnarray}
This set of equations is closed by the Dyson Equation,
Eq.~(\ref{eq:coll.dyson}):
\begin{equation}
G_{ab}^{-1} (\mathbf{k}) = \left[ G_0 (\mathbf{k}) + \Sigma_{\cal G}
(\mathbf{k}) - \Sigma_{\cal F} (\mathbf{k}) \right] \delta_{ab} +
\Sigma_{\cal F} (\mathbf{k}),
\end{equation}
which, in the limit $m \to 1$, according to Eq.~(\ref{eq:coll.p2}), gives:
\begin{equation} \label{eq:coll.SCSAd1}
\mathcal{G}^{-1} (\mathbf{k}) = G_0^{-1} (\mathbf{k}) + \Sigma_{\mathcal{G}}
(\mathbf{k})
\end{equation}
for the diagonal elements, and
\begin{eqnarray} \label{eq:coll.SCSAd2}
\mathcal{F} (\mathbf{k}) &=& - \, \frac{\mathcal{G}^{2} (\mathbf{k})
\Sigma_{\mathcal{F}} (\mathbf{k})}{1 - \mathcal{G} (\mathbf{k})
\Sigma_{\mathcal{F}} (\mathbf{k})}\\
\nonumber
&=& \mathcal{G} (\mathbf{k}) - \frac{1}{\mathcal{G}^{-1} (\mathbf{k})
- \Sigma_{\mathcal{F}} (\mathbf{k})},
\end{eqnarray}
for the off-diagonal ones.

Within the self-consistent screening approximation, the free energy assumes
the following expression~\cite{bray,mezard1,wolynes,wolynes1}:
\begin{equation}
\frac{F(m)}{2 m T} = \textrm{Tr} \ln G^{-1} + \textrm{Tr} \ln D^{-1}
+ \textrm{Tr} \, G \Sigma.
\end{equation}
The evaluation of the traces in the replica space is straightforward.
After that we are able to perform the derivative with respect to the number
of replicas in order to compute the complexity, $\Sigma$, according to
Eq.~(\ref{eq:coll.sconf}). To this aim we have to take the analytical
continuation to $m \to 1$.
The configurational entropy can be written
as the sum of two contributions~\cite{wolynes,wolynes1}:
\begin{equation}
\Sigma = \Sigma^{(1)} + \Sigma^{(2)},
\end{equation}
where
\begin{equation} \label{eq:coll.sc1}
\Sigma^{(1)} = \frac{1}{2} \int \frac{\drv^3 \mathbf{k}}
{(2 \pi)^3} \, \left\{ \ln \left( 1 - \frac{\mathcal{F} (\mathbf{k})}
{\mathcal{G} (\mathbf{k})} \right) + \frac{\mathcal{F} (\mathbf{k})}
{\mathcal{G} (\mathbf{k})} \right\},
\end{equation}
and
\begin{eqnarray} \label{eq:coll.sc2}
\nonumber
\Sigma^{(2)} &=& \frac{1}{2} \int \frac{\drv^3 \mathbf{k}}
{(2 \pi)^3} \, \bigg\{ \ln \left( 1 -
\frac{g T \, \Pi_{\mathcal{F}} (\mathbf{k})}
{1 + g T \, \Pi_{\mathcal{G}} (\mathbf{k})} \right) \\
&& \qquad + \, \frac{g T \, \Pi_{\mathcal{F}} (\mathbf{k})}
{1 + g T \, \Pi_{\mathcal{G}} (\mathbf{k})} \bigg\}.
\end{eqnarray}
From Eqs.~(\ref{eq:coll.sc1}-\ref{eq:coll.sc2}) and
Eq.~(\ref{eq:coll.SCSAd2}), it immediately follows that
$\Sigma = 0$ if $\mathcal{F} (\mathbf{k})$ vanishes or,
equivalently, if $\Sigma_{\mathcal{F}} (\mathbf{k})$ vanishes.

Solving numerically Eqs.~(\ref{eq:coll.SCSAa})-(\ref{eq:coll.SCSAd2})
we get the expressions of the correlators, ${\cal G} ({\bf k})$ and ${\cal F}
({\bf k})$, of the self-energies, $\Sigma_{\cal G} ({\bf k})$ and $\Sigma_{\cal
F} ({\bf k})$, and of the polarization functions, $\Pi_{\cal G} ({\bf k})$
and $\Pi_{\cal F} ({\bf k})$.
Then, according to Eqs.~(\ref{eq:coll.sc1}) and (\ref{eq:coll.sc2}) we
are able to compute the complexity, $\Sigma$.


\begin{thebibliography}{40}
\bibitem{lattice} The model can be thought as the continuum limit of a lattice
gas model described by the following Hamiltonian: ${\cal H} = - \epsilon \sum_{
\langle i j \rangle} n_i n_j + W \sum_{i \ne j} \frac{e^{-|i-j|/\lambda}}{|i-j|} n_i n_j + \mu \sum_i
n_i$, where $n_i=1,0$ whether the site $i$ is occupied by a particle or not. In this 
paper we only consider the values of the chemical potential $\mu$ for which $\langle n_i
\rangle = 0.5$.
\bibitem{nota_scsa} The 
applicability of the SCSA is supported by the fact that the
large $N$ limit is not singular.
\end{thebibliography}
\end{document}